\documentclass[lettersize,journal]{IEEEtran}
\usepackage[HTML]{xcolor}
\definecolor{myred}{HTML}{C00000}
\usepackage[colorlinks=true, linkcolor=myred, citecolor=blue, urlcolor=black]{hyperref}
\usepackage{xspace}
\newcommand{\sysname}{CCAegis\xspace}
\newcommand{\appname}{CCApp\xspace}
\newcommand{\appnames}{CCApps\xspace}
\usepackage{amsmath}
\usepackage{cleveref}
\crefformat{section}{#2§#1#3}
\crefname{section}{§}{§§}
\usepackage{pifont}
\usepackage{adjustbox}
\usepackage{multirow}
\usepackage{amssymb}
\usepackage[ruled,linesnumbered]{algorithm2e}

\begin{document}
\title{More Granular, Less Trust: Enforcing Intra-Process Isolation with Arm CCA in an Untrusted Management Environment}
\author{Shiqi~Liu,
    Zhouqi~Jiang,
    Jie~Wang,
    Wei~Zhou,
    Kun~Sun,
    Zhaohui~Chen,
    and Yulai~Xie
    \thanks{This work was supported in part by the National Key Research and Development Program of China (No.2022YFB4501300), in part by the National Natural Science Foundation of China (No.62202194 and No.62202188). The associate editor coordinating the review of this article and approving it for publication was Prof. Fengwei Zhang. \emph{(Shiqi~Liu and Zhouqi~Jiang contributed equally to this work.} \emph{Corresponding author: Jie~Wang and Wei~Zhou.)}}

    \thanks{Shiqi~Liu, Zhouqi~Jiang, Jie~Wang, and Wei~Zhou are with Hubei Key Laboratory of Distributed System Security, Hubei Engineering Research Center on Big Data Security, School of Cyber Science and Engineering, Huazhong University of Science and Technology, Wuhan, 430074, China, and also with the JinYinHu Laboratory, Wuhan, 430040, China (e-mail: shiqiliu@hust.edu.cn; luojia@hust.edu.cn; wangjie\_s@hust.edu.cn; weizhou\_sec@hust.edu.cn).}

    \thanks{Kun~Sun is with the Center for Secure Information Systems, George Mason University, Fairfax, VA, 22030, USA (e-mail: ksun3@gmu.edu).}

    \thanks{Zhaohui~Chen is with School of Integrated Circuits, Peking University, Beijing, 100871, China, and also with the DAMO Academy, Alibaba Group, Beijing, 100020, China (e-mail: czh@pku.edu.cn).}

    \thanks{Yulai~Xie is with Key Laboratory of Information Storage Systems, Ministry of Education, School of Cyber Science and Engineering, Huazhong University of Science and Technology, Wuhan, 430074, China (e-mail: ylxie@hust.edu.cn).}
    }

\markboth{IEEE TRANSACTIONS ON INFORMATION FORENSICS AND SECURITY}
{LIU \MakeLowercase{et al.}: More Granular, Less Trust: Enforcing Intra-Process Isolation with Arm CCA in an Untrusted Management Environment}

\maketitle

\begin{abstract}
With the increasing adoption of confidential computing, security-sensitive applications are often deployed in confidential virtual machines (CVMs), which reduce reliance on third-party cloud providers. However, privilege attacks originating from the OS remain a significant threat in these environments. Existing finer-grained isolation schemes, such as \textsc{Shelter}, provide process-level protection but are still vulnerable to intra-process attacks and potential collusion between the OS and intra-process adversaries. Many current intra-process isolation techniques continue to depend on the OS to manage and enforce isolation domains, leading to a large Trusted Computing Base (TCB). This gap highlights the need for more granular, less trust-dependent confidential computing solutions.

In this paper, we present \sysname, a system that extends the Arm Confidential Compute Architecture (CCA) to enforce intra-process isolation of sensitive data and operations, safeguarding them from both intra-process adversaries and the OS. We employ static analysis to track the flow of sensitive data and identify functions that handle such data. Permission-switching instructions are inserted at the function call and return points, adjusting permissions via the Granule Protection Table (GPT) to ensure that only designated functions can access the isolated data. Notably, \sysname places trust solely in the Secure Monitor, which configures the GPTs and manages domain switching, thereby minimizing the TCB. We implemented \sysname on both an official emulator and a real development board to assess its performance. Our experimental results show that \sysname effectively isolates sensitive data and operations, with performance overheads ranging from 1.01$\times$ to 1.43$\times$ compared to the original version across real-world cryptographic workloads.
\end{abstract}


\IEEEpeerreviewmaketitle

\maketitle

\section{Introduction}

\IEEEPARstart{P}{rotecting} sensitive data and operations from memory leakage attacks is crucial, particularly when adversaries gain control over privileged software (e.g., OS and hypervisor), which can directly access user-space address space. Recently, confidential computing has gained significant traction, with major chip manufacturers (e.g., AMD, Intel, and Arm) adopting a virtual machine-based isolation model~\cite{amd2023sev,intel2023tdx,arm2023cca}. This model allows users to deploy security-sensitive applications in confidential virtual machines (CVMs) without needing to trust the hypervisor of the cloud service provider. While this approach simplifies the integration of OS services for security-sensitive applications, it remains vulnerable to privilege escalation attacks~\cite{lin2022dirtycred} against the OS—especially when malicious applications share the same CVM.

Arm has enhanced its TrustZone architecture~\cite{alves2004trustzone} with the Confidential Compute Architecture (CCA)~\cite{arm2023cca} in Armv9-A, introducing two new execution environments: the realm world and the root world, enabled by the Realm Management Extension (RME)~\cite{arm2021rme} hardware primitive. This advancement allows third-party developers to seamlessly deploy CVMs within the realm world. 
However, the usability comes at the cost of a large Trusted Computing Base (TCB), which encompasses the guest OS and other privileged software stacks within the realm world and the root world. 

A fundamental limitation of current approaches is their overly coarse-grained isolation, given that only a small fraction of an application’s code is security-sensitive. For example, RContainer~\cite{zhourcontainer} excludes the native OS from its TCB but only extends Arm CCA with container-level isolation. \textsc{Shelter}~\cite{zhang2023shelter} implements isolation at a finer granularity (process-level); however, its monolithic isolation model proves insufficient for defending against intra-process attacks (e.g., buffer overflows~\cite{durumeric2014matter}). Another critical issue is that current intra-process isolation solutions either introduce an excessively large TCB or face significant compatibility challenges. Many solutions predominantly rely on OS-managed mechanisms, making them vulnerable to privilege escalation attacks~\cite{lin2022dirtycred}. For instance, libmpk~\cite{park2019libmpk} and ERIM~\cite{vahldiek2019erim} leverage Memory Protection Keys (MPK) for user-space domain switching, but domain configuration still involves OS intervention. 
On Arm, Shreds~\cite{chen2016shreds} and ARMLock~\cite{zhou2014armlock} build compartments atop legacy memory domains; they likewise depend on the OS for enforcement, and memory domains are not supported in AArch64 and are deprecated on modern Arm platforms, limiting their applicability.
Other approaches~\cite{park2020nested,gu2022hardware} use Intel SGX~\cite{intel2023sgx} to protect against privilege escalation and implement fine-grained isolation by creating multiple domains within a single enclave. However, these approaches require hardware modifications, which lead to compatibility issues with Arm architectures.

The key to addressing these issues lies in exploring a new mechanism that \emph{\textbf{leverages off-the-shelf hardware features on Arm platforms to achieve intra-process isolation within an untrusted management environment}}. Beyond the usual virtual address partitioning via page tables, Arm CCA provides the Granule Protection Table (GPT) to split a process’s physical address space into protection domains. Crucially, GPT configuration and switching are performed by the Secure Monitor in the Root world (trusted firmware), which removes the OS from the TCB and keeps it minimal. Because GPT enforces permissions directly over physical memory under the Monitor’s control, \textit{per-module} vetting of dynamically loaded kernel code (as required by same-privilege schemes such as Nested Kernel~\cite{dautenhahn2015nested}) is not needed to prevent isolation-breaking instructions.
Building on these observations, we propose \sysname, a system that enforces fine-grained, low-trust isolation using the GPT. \sysname allocates sensitive intra-process data to a dedicated physical memory region and isolates this data by creating separate GPTs for sensitive and non-sensitive code. Specifically, only the GPT associated with sensitive code is configured to grant access to the protected memory region. \sysname achieves domain switching by switching GPTs at the boundaries of sensitive code.

However, directly employing GPT for intra-process isolation faces two challenges. First, existing isolation models—whether native VM-level isolation or subsequent improvements enabling container/process-granular isolation~\cite{zhourcontainer,zhang2023shelter}—are monolithic by design, offering deployment friendliness that requires no restructuring of application logic. In contrast, \emph{\textbf{extending CCA to achieve intra-process isolation imposes significant porting costs}}, as it requires developers to manually identify sensitive code boundaries and make invasive modifications to application code (e.g., inserting GPT-switching instructions).
Second, since the process management environment is provided by the untrusted OS, \emph{\textbf{the isolation enforced by the Monitor introduces a semantic gap between the user space and the root world}}. Specifically, the process operates within a virtual address space, while the Monitor interacts directly with physical memory addresses, lacking memory mapping information for the pages containing sensitive data.

To maintain the deployment-friendly nature of CCA, \sysname modifies the LLVM compiler to perform static analysis (including points-to and taint analysis) on security-sensitive application source code, tracking the propagation of sensitive data (e.g., cryptographic keys) and identifying the minimal subset of functions that access such data (\cref{sec:static_analysis}). \sysname then inserts GPT-switching instructions at the identified code boundaries to enforce memory isolation. Additionally, given the untrusted OS, \sysname automatically replaces system calls related to the memory allocation and usage of sensitive data.

To bridge the semantic gap with user space, \sysname introduces a kernel driver that initializes and maps virtual memory for sensitive data, as well as creates shadow page tables (parallel to native page tables for direct OS updates) for protected processes (\cref{sec:isolation}). The root world ensures isolation against the OS by marking the physical pages containing sensitive data and the shadow page tables as inaccessible in the OS's GPT, preventing unauthorized access to sensitive data or tampering with memory mappings. Notably, this driver is not considered trusted to maintain a minimal TCB. Therefore, \sysname's root world component must verify the correctness of memory mappings by inspecting the shadow page tables.

We implement \sysname on an official Arm emulator and a real Arm hardware SoC.
The code size of \sysname is around 5K lines of code (LoCs), comprising 2K LoCs for static analysis modules in C++/Rust and 3K LoCs for runtime isolation modules in C. 
We evaluate the performance of \sysname across five widely used real-world applications (e.g., \texttt{Nginx}). The evaluation shows that \sysname achieves strong isolation by protecting only the sensitive parts of a program containing 10.57\% to 35.26\% of the total LoCs, with a moderate performance overhead ranging from 1.01$\times$ to 1.43$\times$ compared to the original version. 
We release the source code for \sysname at \url{https://github.com/erhade/CCAegis}.

In summary, we make the following contributions: 
\begin{itemize}
    \item We present \sysname, an intra-process isolation system built on Arm CCA using the GPT primitive, designed to prevent sensitive data leakage in an untrusted management environment.
    \item We address two key challenges: automating program partitioning and isolation instrumentation through static analysis, ensuring \sysname remains deployment-friendly, and bridging the semantic gap via an untrusted kernel driver and a trusted Monitor, all while maintaining a minimal TCB.
    \item We evaluate \sysname on both an official emulator and a real development board. The results show that \sysname effectively protects sensitive data with moderate performance overhead.
\end{itemize}

\section{Background} \label{sec:background}

\subsection{Arm Confidential Computing Architecture} \label{subsec:cca}

\begin{figure}[!h]
  \centering
  \includegraphics[width=3.3 in]{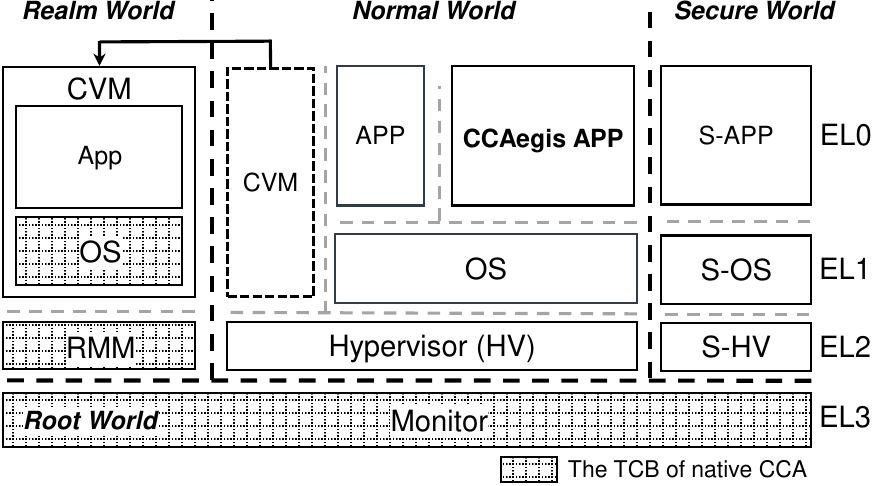}
  \caption{The architecture of CCA.}
  \label{fig:cca}
\end{figure}

\begin{table}[!h]
  \caption{Physical address access permissions of Arm CCA.}
  \label{tab:permission}
  \centering
  \begin{adjustbox}{max width=\linewidth}
  \begin{tabular}{ccccc}
  \hline

  \hline
  \textbf{Security State} & \textbf{Normal World} & \textbf{Realm World} & \textbf{Secure World} & \textbf{Root World} \\
  \hline
    \emph{Normal} & \ding{51} & \textcolor[HTML]{c00000}{\ding{55}} & \textcolor[HTML]{c00000}{\ding{55}} & \textcolor[HTML]{c00000}{\ding{55}}\\
    \emph{Realm} & \ding{51} & \ding{51} & \textcolor[HTML]{c00000}{\ding{55}} & \textcolor[HTML]{c00000}{\ding{55}}\\
    \emph{Secure} & \ding{51} & \textcolor[HTML]{c00000}{\ding{55}} & \ding{51} & \textcolor[HTML]{c00000}{\ding{55}}\\
    \emph{Root} & \ding{51} & \ding{51} & \ding{51} & \ding{51}\\
  \hline

  \hline
  \end{tabular}
  \end{adjustbox}
\end{table}

\begin{table*}[!t]
    \caption{Comparison of \sysname with related TEE systems. 
    \ding{61} CURE supports resilient TEE; here, we consider its finest granularity.}
  \label{tab:comparison}
  \centering
  \begin{adjustbox}{max width=\linewidth}
\begin{tabular}{rcccccccc}
  \hline

  \hline
\textbf{System} & \textbf{Isolation} & \textbf{Isolation} & \textbf{Trusted Computing} & \textbf{Intra-Process} & \textbf{Deployment} & \textbf{Arm} & \textbf{Unrestricted} \\
\textbf{Name} & \textbf{Primitive} & \textbf{Granularity} & \textbf{Base (LoCs)} & \textbf{Isolation} & \textbf{Friendliness} & \textbf{Support} & \textbf{Data Volume} \\
 \hline
TDX~\cite{intel2023tdx} & TDX & VM & TDX Module (46K) + OS in Trust Domain ($\ge$25M) & \textcolor[HTML]{c00000}{\ding{55}} & \ding{51} & \textcolor[HTML]{c00000}{\ding{55}} & \ding{51} \\
 
SGX~\cite{intel2023sgx} & SGX & Part of App & - & \ding{51} & \textcolor[HTML]{c00000}{\ding{55}} & \textcolor[HTML]{c00000}{\ding{55}} & \ding{51} \\

Glamdring~\cite{lind2017glamdring} & SGX & Function & - & \ding{51} & \ding{51} & \textcolor[HTML]{c00000}{\ding{55}} & \ding{51} \\

SeCage~\cite{liu2015thwarting} & VMFUNC & Function & KVM Hypervisor (10M) & \ding{51} & \ding{51} & \textcolor[HTML]{c00000}{\ding{55}} & \ding{51} \\

SEV (-ES, -SNP)~\cite{amd2023sev} & SEV & VM & Firmware (N/A) + OS in Confidential VM ($\ge$25M) & \textcolor[HTML]{c00000}{\ding{55}} & \ding{51} & \textcolor[HTML]{c00000}{\ding{55}} & \ding{51} \\

Keystone~\cite{lee2020keystone} & PMP & Part of App & Monitor (37K) + Runtime (9K) & \ding{51} & \textcolor[HTML]{c00000}{\ding{55}} & \textcolor[HTML]{c00000}{\ding{55}} & \ding{51} \\

CURE~\cite{bahmani2021cure} & Customized HW & Part of App\textsuperscript{\ding{61}} & Monitor (3K) + Runtime (16M, Opt.) & \ding{51} & \textcolor[HTML]{c00000}{\ding{55}} & \textcolor[HTML]{c00000}{\ding{55}} & \ding{51} \\

Sanctum~\cite{costan2016sanctum} & Customized HW & Part of App & Monitor (2K) + glibc (0.9M) & \ding{51} & \textcolor[HTML]{c00000}{\ding{55}} & \textcolor[HTML]{c00000}{\ding{55}} & \ding{51} \\

\textsc{PengLai}~\cite{feng2021scalable} & Customized HW & Part of App & Monitor (43K) + musl libc (80K) & \ding{51} & \textcolor[HTML]{c00000}{\ding{55}} & \textcolor[HTML]{c00000}{\ding{55}} & \ding{51} \\

TrustZone~\cite{alves2004trustzone} & TZASC & Part of App & Monitor (0.5M) + OP-TEE (0.4M) & \ding{51} & \textcolor[HTML]{c00000}{\ding{55}} & \ding{51} & \ding{51} \\

Sanctuary~\cite{brasser2019sanctuary} & TZASC & App & Monitor (0.5M) + Runtime (0.5M) & \textcolor[HTML]{c00000}{\ding{55}} & \ding{51} & \ding{51} & \ding{51} \\

CaSE~\cite{zhang2016case} & TZASC & App & Trusted Boot (500) + Controller (500) & \textcolor[HTML]{c00000}{\ding{55}} & \ding{51} & \ding{51} & \textcolor[HTML]{c00000}{\ding{55}} \\

Ginseng~\cite{yun2019ginseng} & EL3/Virtualization & Function & Monitor (0.5M) & \ding{51} & \ding{51} & \ding{51} & \textcolor[HTML]{c00000}{\ding{55}} \\

CCA~\cite{arm2023cca} & CCA GPC & VM & Monitor (0.5M) + RMM (33K) + OS in CVM ($\ge$26M) & \textcolor[HTML]{c00000}{\ding{55}} & \ding{51} & \ding{51} & \ding{51} \\

ACAI~\cite{sridhara2024acai} & CCA GPC & VM & Monitor (0.5M) + RMM (23K) + OS in CVM ($\ge$26M) & \textcolor[HTML]{c00000}{\ding{55}} & \ding{51} & \ding{51} & \ding{51} \\

CAGE~\cite{wang2024cage} & CCA GPC & VM & Monitor (0.5M) + RMM (26K) + OS in CVM ($\ge$26M) & \textcolor[HTML]{c00000}{\ding{55}} & \ding{51} & \ding{51} & \ding{51} \\

RContainer~\cite{zhourcontainer} & CCA GPC & Container & Monitor (0.5M) + Mini-OS (5K) & \textcolor[HTML]{c00000}{\ding{55}} & \ding{51} & \ding{51} & \ding{51} \\

\textsc{Shelter}~\cite{zhang2023shelter} & CCA GPC & App & Monitor (0.5M) & \textcolor[HTML]{c00000}{\ding{55}} & \ding{51} & \ding{51} & \ding{51} \\

\hline
\textbf{\sysname} & \textbf{CCA GPC} & \textbf{Function} & Monitor (0.5M) & \ding{51} & \ding{51} & \ding{51} & \ding{51} \\
  \hline

  \hline
\end{tabular}
\end{adjustbox}
\end{table*}

In the latest Armv9.2-A, Arm introduced the Confidential Compute Architecture (CCA)~\cite{arm2023cca} along with a core hardware support feature known as the Realm Management Extension (RME)~\cite{arm2021rme}, as depicted in \autoref{fig:cca}. CCA introduces the realm world and the root world in addition to the existing TrustZone architecture. 
The TCB of CCA encompasses the OS within confidential virtual machines (CVMs) running in the realm world, the Realm Management Monitor (RMM) which is responsible for managing and isolating these CVMs, and the Monitor operating in the root world. RME introduces new data structures, such as the Granule Protection Table (GPT), and new hardware mechanisms, such as the Granule Protection Check (GPC), to enable fine-grained access control over physical memory. The GPT tracks the physical pages assigned to each world. Each entry in the GPT can be configured with one of six attributes, determining its association with a specific world—\emph{normal}, \emph{realm}, \emph{secure}, or \emph{root}—or set to universal accessibility (\emph{all-access}) or complete inaccessibility (\emph{no-access}). Typically, CCA uses a single GPT to manage attributes for all physical memory. In contrast, \sysname constructs multiple GPTs, allowing different entities to enforce distinct access permissions over the same physical memory, thereby isolating specific memory regions. During each memory access, the CPU retrieves the attributes of the current memory page from the GPT and performs a GPC, as defined in \autoref{tab:permission}, to verify the access's legitimacy. If the access is deemed unauthorized, a Granule Protection Fault (GPF) is immediately triggered.

\subsection{Trust Firmware-A} \label{subsec:tfa}
Trust Firmware-A (TF-A)~\cite{arm2024tfa} is a reference firmware implementation for Arm-based processors, responsible for initializing hardware and ensuring a secure boot process for CCA. Unlike TrustZone, which separates Exception Level 3 (EL3) into the secure world, CCA uses EL3 to establish the root world. TF-A operates in the root world and manages the isolation of the four worlds. With the highest privilege level, it prevents other worlds from modifying the configuration of the GPT and GPC. Given that TF-A (590K LoCs) is significantly smaller than the Linux kernel (26M LoCs), using it as the supervisory layer results in a smaller TCB.

\section{Motivation and Overview} \label{sec:overview}

\subsection{Motivation} \label{subsec:motivation}

Trusted Execution Environments (TEEs) can safeguard sensitive code and data without relying on trust in privileged software, and many approaches have explored isolation designs at varying granularities, as shown in \autoref{tab:comparison}. However, they still face the following critical limitations:

\noindent \textbf{Excessively Large TCB.}
Both native confidential computing solutions—including AMD's SEV~\cite{amd2023sev}, Intel's TDX~\cite{intel2023tdx}, and Arm's CCA~\cite{arm2023cca}—as well as extended proposals like ACAI~\cite{sridhara2024acai} and CAGE~\cite{wang2024cage} (which aim to address gaps in support for accelerators like GPUs) adopt VM-level isolation. However, these solutions require trust in the OS within the CVM, introducing millions of LoCs to the TCB when using the Linux kernel. In contrast, Sanctuary~\cite{brasser2019sanctuary} uses the Zircon microkernel as the runtime for security-sensitive applications, though it still expands the TCB by approximately 500K LoCs. Notably, CCA further requires trust in the RMM to manage CVMs, adding over 20K LoCs to the TCB.

\noindent \textbf{Overly Coarse Isolation Granularity.}
Other solutions directly isolate security-sensitive applications without requiring trust in the OS, maintaining a small TCB. For example, CaSE~\cite{zhang2016case} isolates applications within the L2 Cache, while RContainer~\cite{zhourcontainer} and \textsc{Shelter}~\cite{zhang2023shelter} leverage multiple GPTs to prevent the OS from accessing protected container/application. However, their isolation granularity remains overly coarse, leaving them vulnerable to intra-process vulnerabilities~\cite{durumeric2014matter}.

\begin{figure*}[!t]
  \centering
  \includegraphics[width=6.2 in]{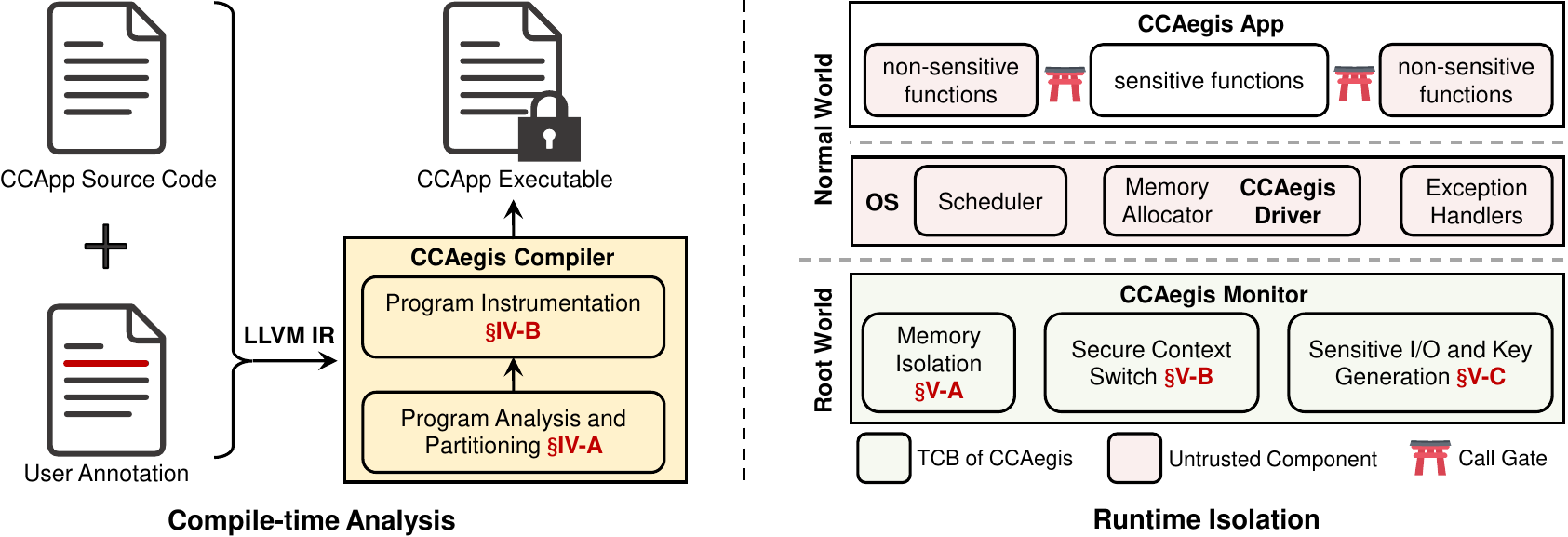}
  \caption{The workflow of \sysname' compile-time analysis and its runtime isolation architecture.}
  \label{fig:workflow}
\end{figure*}

\noindent \textbf{Lack of Deployment/Porting Friendliness.}
Various TEEs provide fine-grained isolation for secure parts of applications across different architectures, including those on Intel~\cite{intel2023sgx}, Arm~\cite{alves2004trustzone}, and RISC-V~\cite{feng2021scalable,lee2020keystone,bahmani2021cure,costan2016sanctum}. However, these TEEs require developers to manually partition programs and adjust isolation interfaces, a process that can be cumbersome and error-prone. To improve deployment friendliness, several solutions~\cite{lind2017glamdring,yun2019ginseng,liu2015thwarting} have developed automated program partitioning using program analysis technologies. 
However, Glamdring~\cite{lind2017glamdring} and SeCage~\cite{liu2015thwarting} rely on Intel-specific hardware features, limiting their portability on Arm platforms. Ginseng~\cite{yun2019ginseng}, which uses registers for isolation, is constrained by the limited data volume it can protect.

In conclusion, most TEEs~\cite{arm2023cca, zhourcontainer, zhang2023shelter} exhibit overly coarse isolation granularity (e.g., VMs, containers, or applications). Even existing fine-grained TEEs face limited practicality due to their heavy reliance on manual developer effort~\cite{keystone2023eyrie, bahmani2021cure} or their use of caches~\cite{zhang2016case} or registers~\cite{yun2019ginseng} as isolation primitives.

\subsection{\sysname Architecture} \label{subsec:architecture}

We designed \sysname to address the limitations of prior approaches, motivated by two key observations. First, manual effort can be minimized by requiring developers to annotate only a small set of initial sensitive data, with taint propagation analysis automatically identifying subsequent sensitive data and enforcing instrumentation-based protection. Second, memory-based isolation is more suitable than cache- or register-centric approaches for protecting general-purpose applications with large data volumes, and the GPT inherently enables memory access control without trusting the OS. Since the Secure Monitor is inevitably in the TCB, a stage-2/Realm-EL2 design would additionally pull the RMM into the TCB, whereas GPT keeps enforcement within the existing firmware boundary. Building on these findings, \sysname operates through two fundamental stages, as depicted in \autoref{fig:workflow}: Compile-time analysis (\cref{sec:static_analysis}) is conducted using the \emph{\sysname-Compiler}, while runtime isolation (\cref{sec:isolation}) is managed by the \emph{\sysname-Monitor} located in the root world.

\noindent \textbf{Compile-time Analysis.}
\sysname's analysis consists of five steps to construct an executable with sensitive function isolation properties. Notably, this analysis is performed \emph{offline}, making it immune to threats from runtime attackers. Initially, developers annotate the program with initial sensitive data and their final forms, establishing sources and sinks for subsequent analysis. Given the source code of a \emph{\sysname-App} (\appname), the \emph{\sysname-Compiler} converts it into LLVM Intermediate Representation (IR) along with its call graph. It then performs points-to and taint analysis to track the propagation of sensitive data, identifying sensitive operations and the functions that contain them (\cref{subsec:Partitioning}). The \emph{\sysname-Compiler} controls access to sensitive data by inserting call gates at the entry and exit points of identified sensitive functions. These gates invoke the \emph{\sysname-Monitor} to grant and revoke access permissions (\cref{subsec:instrumentation}). Finally, the \emph{\sysname-Compiler} completes the compilation and linking processes to produce the executable.

\noindent \textbf{Runtime Isolation.}
Basic capabilities such as scheduling and exception handling are managed by the untrusted OS. To bridge the semantic gap between \appnames and the \emph{\sysname-Monitor}, the driver we introduced allocates and maps virtual memory for sensitive functions. However, critical security operations—such as physical memory delegation and recycling (\cref{subsec:memory}), state transitions between trusted and untrusted components (including transitions within intra-process and between \appnames and the OS) (\cref{subsec:exception}), as well as sensitive I/O operations and key generation (\cref{subsec:io_key})—are overseen and authorized by the \emph{\sysname-Monitor}.

\begin{figure}[!t]
  \centering
  \includegraphics[width=3 in]{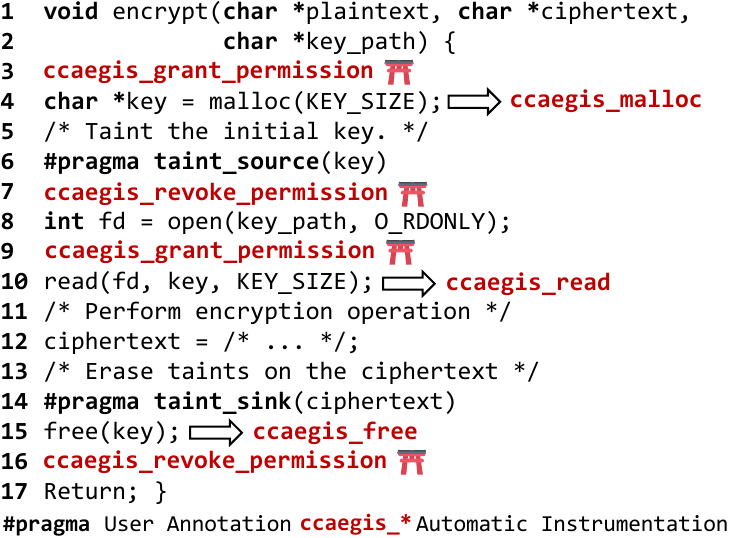}
  \caption{An example of how \sysname automatically generates instrumentation to protect cryptographic programs by leveraging developer annotations.}
  \label{fig:code-example}
\end{figure}

\noindent \textbf{Programming Model.}
Developers only need minimal annotations to utilize \sysname's isolation capabilities, as depicted in \autoref{fig:code-example}. They begin by marking the initial \emph{taint source}, usually the user-defined cryptographic keys (line 6), and designate the point where data becomes ciphertext as the \emph{taint sink} (line 14), indicating where taint propagation should halt. Following the taint propagation analysis, the \emph{\sysname-Compiler} identifies sensitive functions and automates the protection process. To isolate these functions, it inserts call gates at critical boundaries—specifically at the entry (line 3) and exit (line 16) of sensitive functions, and before (line 7) and after (line 9) calls to external non-sensitive functions. Note that \sysname does not require explicit protection for file descriptors, as signature verification is enforced on file contents during read operations. It also replaces unsafe key-related functions with secure functions provided by the \emph{\sysname-Monitor} for memory allocation (line 4), memory release (line 15), and file loading (line 10).

\subsection{Threat Model and Assumptions} \label{subsec:threat_model} 

The TCB of \sysname only comprises the root world Monitor, which is verified through vendor signatures and securely loaded via secure boot. \sysname trusts the hardware, assuming that Arm CCA and RME conform to their specifications. 
Its primary goal is to safeguard the confidentiality and integrity of keys within cryptographic programs, shielding them from intra-process adversaries. Additionally, it protects against privileged adversaries including the OS, hypervisor, and components within both secure and realm worlds. It also enforces mutual isolation, preventing malicious \appnames from attacking other \appnames and privileged software.

\noindent \textbf{Attackers' capabilities.}
For intra-process adversaries, we consider their ability to exploit memory vulnerabilities, such as buffer overflows~\cite{durumeric2014matter}. Additionally, they may execute control flow hijacking~\cite{shacham2007geometry} to bypass the permission controls enforced by the call gates. For privileged adversaries, we account for more advanced key theft methods, including directly reading memory, register states, and I/O data from \appnames. These adversaries may also access key-related memory via Direct Memory Access (DMA) and hijack control flow by modifying register states or altering the \appname's stack. Furthermore, we also focus on memory-oriented Iago vectors that directly impact confidentiality, including forged or aliased mappings, tampered pointers or lengths, and page-lifetime manipulations that could cause secrets to leave protected regions~\cite{checkoway2013iago,cui2021emilia}.

\noindent \textbf{Out of scope.} 
We do not consider scenarios where sensitive functions intentionally leak secrets or security risks within sensitive function code, which could be mitigated by orthogonal techniques, such as software vulnerability detection~\cite{haller2013dowsing}.
Additionally, we exclude hardware attacks such as bus snooping~\cite{lee2020off} and DRAM analysis (e.g., Cold Boot~\cite{yitbarek2017cold} and Rowhammer attacks~\cite{kim2014flipping}). Arm CCA plans to implement the Memory Protection Engine (MPE) for memory encryption and possibly integrity protection, which should mitigate these hardware threats in the future. Denial-of-Service (DoS) attacks and side-channel attacks (e.g., Meltdown~\cite{lipp2018meltdown} and Spectre~\cite{kocher2020spectre}) are also out of scope.

\section{\sysname Compile-time Analysis} \label{sec:static_analysis}

\subsection{Program Analysis and Partitioning} \label{subsec:Partitioning}

\noindent\textbf{Overview.}
We treat the point where user–defined cryptographic secrets enter a \appname (for example, a key read from disk) as a \emph{taint source}, and the locations where data becomes ciphertext or a signature as \emph{taint sinks}; once data is in its final protected form, further tracking is unnecessary. Given sources and sinks, the \emph{\sysname-Compiler} performs points-to and taint analysis~\cite{grech2017p,machiry2017dr} to follow secret propagation, identifies all variables and accesses on the secret path, and then classifies any \texttt{load} or \texttt{store} that touches these addresses as a \emph{crypto operation}. A function is a \emph{crypto function} if it contains at least one crypto operation. Isolation is enforced at function granularity over this discovered set.

\noindent\textbf{Taint IR.}
Because standard compiler IRs such as LLVM IR do not provide native dataflow tracking, we introduce a \emph{Taint IR} layered over the compiler IR. Taint IR (i) normalizes constructs across IRs so that results can be reapplied to the original IR without binding to a specific compiler, (ii) tags instructions with context–sensitive identifiers to make the analysis context aware, and (iii) incorporates \emph{index–based} modeling that records byte–range offsets for structured and array objects. The index model mitigates over–tainting common in classical Andersen–style analysis by tracking structure elements precisely.

\begin{figure*}[!t]
  \centering
  \begin{adjustbox}{max width=\textwidth}
  $\begin{gathered}
  \frac{\mathbf{alloc}\, v, len}{\left(l_v,[0,len)\right)\in v_{pto}}\,\text{Alloc}\quad
  \frac{\mathbf{load}\, p,q\; (l_r,i)\in q_{pto}\; (l_x,j)\in r_{pto}}{(l_x,\mathrm{INDEX}(j,i))\in p_{pto}}\,\text{Load}\quad
  \frac{\mathbf{store}\, p,q\; (l_r,i)\in p_{pto}\; (l_x,j)\in q_{pto}}{(l_x,j)\in r_{pto}}\,\text{Store}\\[0.8ex]
  \frac{\mathbf{index}\, p,q,i\; (l_x,j)\in q_{pto}}{(l_x,\mathrm{INDEX}(j,i))\in p_{pto}}\,\text{Index}\quad
  \frac{\mathbf{merge}\, p,\{q_1,\dots,q_n\}\; (l_x,i)\in q_{k,pto}}{(l_x,i)\in p_{pto}}\,\text{Merge}\quad
\frac{\displaystyle{
  \genfrac{}{}{0pt}{}{%
    c:\mathbf{call}\, f(a_1,\dots,a_n)\!\to\! b,\; c':f(p_1,\dots,p_n)\!\to\! r%
  }{%
    (l_x,i)\in c:a_{k,pto}\;\; (l_y,j)\in c':r_{pto}%
  }%
}}{%
  (l_x,i)\in c':p_{k,pto}\;\; (l_y,j)\in c:b_{pto}%
}\,\text{CallRet}
  \end{gathered}$%
  \end{adjustbox}
  \caption{Index assisted points-to analysis constraints in Taint IR.}
  \label{fig:points-to}
\end{figure*}

For any global or local variable $v$ within a function, we define its corresponding \textbf{points-to analysis instructions}, denoted as $v_{pto}$, as the set of tuples $\{(l_v, i)\}$. Here, $l_v$ represents the memory location it point to, and $i$ is its containing member variables as byte range index. 
The points-to analysis IR set includes the following instructions:

\begin{enumerate}
    \item $\mathbf{alloc}\,\,v, len$. This instruction creates a memory location $l_v$ pointed to by variable $v$ with memory size $len$. The value of $len$ is related to the memory representation, which we assume remains unchanged within one module during analysis. 
    \item $\mathbf{load}\,\,p, q$. This instruction reads the memory location pointed to by address variable $q$ and writes its contents into $p$.
    \item $\mathbf{store}\,\,p, q$. This instruction writes the contents of variable $q$ into the memory area pointed to by address variable $p$. 
    \item $\mathbf{index}\,\,p, q, i$. This instruction extracts the memory location pointed to address $q$ with index range parameter $i$, and writes location into result address $p$. 
    \item $\mathbf{merge}\,\,p, \{q_1, ..., q_n\}$. This instruction merges the contents of variables $q_1, ..., q_n$ into result variable $p$ according to specific rules, usually corresponding to unary and binary operations or bit conversions in compiler IR. 
    \item $\mathbf{call}\,\,f(a_1, ..., a_n) \rightarrow b$. This instruction calls function $f$ with arguments $a_1, ..., a_n$ and writes the return value into variable $b$.
    \item $\mathbf{return}\,\,r$. This instruction uses variable $r$ as the return value and terminates the current function's execution.
\end{enumerate}

From these instructions we derive index aware points to constraints (see \autoref{fig:points-to}). Taint analysis results are derived from points-to analysis results. To incorporate source annotations, the Taint IR adds two \textbf{taint analysis instructions}:

\begin{enumerate}
\item $\mathbf{taint}\,\,p$. This instruction marks the memory location pointed to by $p$ as tainted. Note that the points-to set $p_{pto}$ includes memory locations and their corresponding indices. This instruction only taints the memory location within the specified index range, but not beyond it.
\item $\mathbf{erase}\,\,p$. This instruction erases the taint paths associated with memory locations and indices pointed by $p$.
\end{enumerate}

According to the above definitions, we produce Taint IR, which are converted from compiler IRs. Note that Taint IR and compiler IRs do not necessarily correspond one-to-one. For instance, the $\mathbf{index}$ instruction determines the $\textit{i}$ parameter from a list of indices and the module-level memory layout in one compiler IR instruction.

\begin{algorithm} [!t]
  \caption{Taint Analysis Algorithm}\label{alg:taint-analysis}
  \SetAlgoLined
  \SetKwFunction{concat}{concat}
  \SetKwFunction{makeseq}{makeseq}
  \KwData{Points-to sets $\{v_{pto}\}$, taint analysis result $\{v_{taint}\}$, current Taint IR instruction $I$}
  \KwResult{Modified taint analysis result $\{v_{taint}\}$}
  \Switch{type of I}{
    \uCase{load p, q}{
      \ForAll{$(l_r, i) \in q_{pto}$, $(l_x, j) \in r_{pto}$, $(s, k) \in x_{taint}$}{
        $p_{taint} \leftarrow p_{taint} \cup \{(\concat(s, I), j \cap k)\}$
      }
    }
    \uCase{store p, q}{
      \ForAll{$(l_r, i) \in p_{pto}$, $(l_x, j) \in q_{pto}$, $(s, k) \in x_{taint}$}{
        $r_{taint} \leftarrow r_{taint} \cup \{(\concat(s, I), i)\}$
      }
    }
    \uCase{index p, q, i}{
      \ForAll{$(s, j) \in q_{taint}$}{
        $p_{taint} \leftarrow p_{taint} \cup \{(\concat(s, I), i \cap j)\}$
      }
    }
    \uCase{merge $p, \{q_1, ..., q_n\}$}{
      \ForAll{$q_k \in \{q_1, ..., q_n\}$, $(s, i) \in q_{k,taint}$}{
        $p_{taint} \leftarrow p_{taint} \cup \{(\concat(s, I), i)\}$
      }
    }
    \uCase{taint p}{
      \lForAll{$(l_r, i) \in p_{pto}$}{
        $r_{taint} \leftarrow \{(\makeseq(I), i)\}$
      }
    }
    \uCase{erase p}{
      \lForAll{$(l_r, i) \in p_{pto}$}{
        $r_{taint} \leftarrow \varnothing$
      }
    }
    \uCase{call f$(a_1, ..., a_n)\rightarrow b$}{
      $c':p_{k,taint} \leftarrow c:a_{k,taint}$
    }
    \uCase{return r}{
      $c:b_{taint} \leftarrow c':r_{taint}$ , where $b$ is the return variable in corresponding $call$ instruction
    }
  }
\end{algorithm}

\noindent\textbf{Taint Analysis Algorithm.}
We detail our taint analysis algorithm in \autoref{alg:taint-analysis}.
The algorithm starts at an entry function linearly scan our transferred point-to and Taint IR instructions of program. 
Note that
the different locations call instructions 
correspond to distinct contexts $c$, making our analysis algorithm context-sensitive.
Finally, we give a definition of taint analysis results for each variable. The \textbf{taint analysis result} for a variable $v$ is defined as $v_{taint}=\{(s, k)\}$, comprising pairs of the taint propagation flow $s$ and the corresponding memory location index $k$. Here, the taint propagation flow $s=(I_1, I_2, ..., I_n)$ represents the sequence of all Taint IR instructions $I$ involved in the propagation chain of $v$.
The taint analysis results will encompass all tainted local and global variables and memory locations. These can then be utilized as sensitive variables and operations in the following instrumentation process.

\subsection{Program Instrumentation} \label{subsec:instrumentation}

Instrumentation for security-sensitive applications must enforce two critical guarantees. First, only crypto functions are granted access to protected memory regions. Second, secure interactions between crypto functions and non-crypto contexts must be ensured, including:
(1) Access to non-sensitive global variables,
(2) Invocation of non-crypto functions within \appnames,
(3) Execution of untrusted system calls.

\noindent \textbf{Secure Memory Management.} 
Given that the minimum isolation granularity of the GPT is 4KB, \sysname optimizes memory usage by reallocating cryptographic keys to a contiguous crypto buffer. These cryptographic keys are typically scattered across a process's memory, residing in the stack, heap, and global variables. To manage these keys efficiently, the \emph{\sysname-Monitor} establishes the crypto stack, crypto heap, and crypto data segment—each contained within the crypto buffer. To secure access to \emph{non-sensitive} global variables from crypto functions, the \emph{\sysname-Compiler} copies the values of these global variables into the crypto stack. It then verifies and accesses these values directly from the crypto stack.

\begin{figure}[!htbp]
  \centering
  \includegraphics[width=3.5 in]{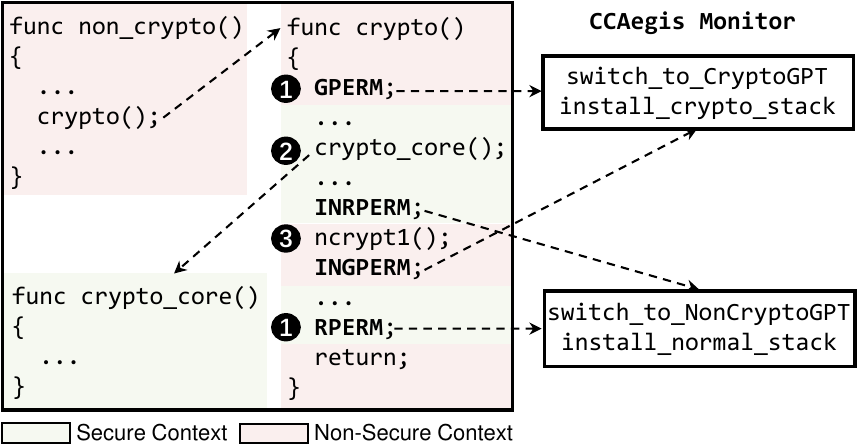}
  \caption{Execution flow through call gate instrumentation.}
  \label{fig:instrument}
\end{figure}

\noindent \textbf{Call Gate Generation.}
The \emph{\sysname-Compiler} isolates crypto functions by strategically inserting call gates. When entering a crypto function, the \emph{\sysname-Monitor} switches to the Crypto GPT to grant access to the crypto buffer, and swaps out the normal stack for a crypto stack. Conversely, upon leaving the crypto function, the \emph{\sysname-Monitor} switches to the Non-Crypto GPT to revoke access, reverts to the original normal stack, and saves the crypto stack pointer.
This process is detailed in three steps, as depicted in \autoref{fig:instrument}:
\ding{182} \emph{Function Entry and Exit:} For each crypto function, the \emph{\sysname-Compiler} inserts a \underline{G}rant \underline{Perm}ission (\texttt{GPERM}) call gate at the function entry and a \underline{R}evoke \underline{Perm}ission (\texttt{RPERM}) call gate before the function returns to ensure the function operates within a secure context.
\ding{183} \emph{Nested Crypto Functions:} If a crypto function is called by another crypto function, additional call gates at the entry and exit of the callee are unnecessary, as the secure context is already established by the caller.
\ding{184} \emph{Interactions with Non-Crypto Functions:} For non-crypto calls within a crypto function, the compiler inserts an \underline{In}ternal \underline{R}evoke \underline{Perm}ission (\texttt{INRPERM}) call gate before the call to isolate the context and an \underline{In}ternal \underline{G}rant \underline{Perm}ission (\texttt{INGPERM}) call gate after the call to re-establish the secure context.

\begin{figure}[!h]
  \centering
  \includegraphics[width=3.5 in]{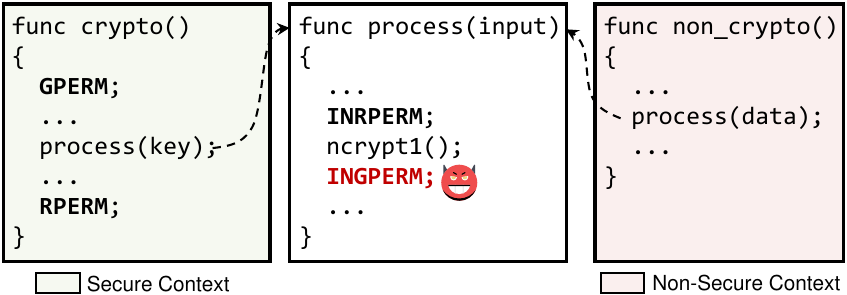}
  \caption{Call gate exploitation when called by different contexts.}
  \label{fig:context}
\end{figure}

\noindent \textbf{Context-Sensitive Isolation.} 
A function may be identified as either a crypto or non-crypto function depending on the context in which it is called, as illustrated in \autoref{fig:context}. For example, when called by \texttt{crypto()} with a key as an argument, \texttt{process()} functions as a crypto function. In this case, inserting call gates at the entry or exit of \texttt{process()} would be redundant. However, if \texttt{process()} contains non-crypto functions, it still requires \texttt{INRPERM} and \texttt{INGPERM} to secure the crypto context. Conversely, when \texttt{non\_crypto()} calls \texttt{process()}, the untrusted \texttt{non\_crypto()} function could exploit the \texttt{INGPERM} instruction within \texttt{process()} to gain access to the crypto buffer. To mitigate this risk, the \emph{\sysname-Compiler} uses context-sensitive analysis to generate different versions of \texttt{process()}. This ensures that only the version invoked within a crypto context is treated as a crypto function, and it is properly fortified with call gates.

\begin{figure}[!htbp]
  \centering
  \includegraphics[width=3.5 in]{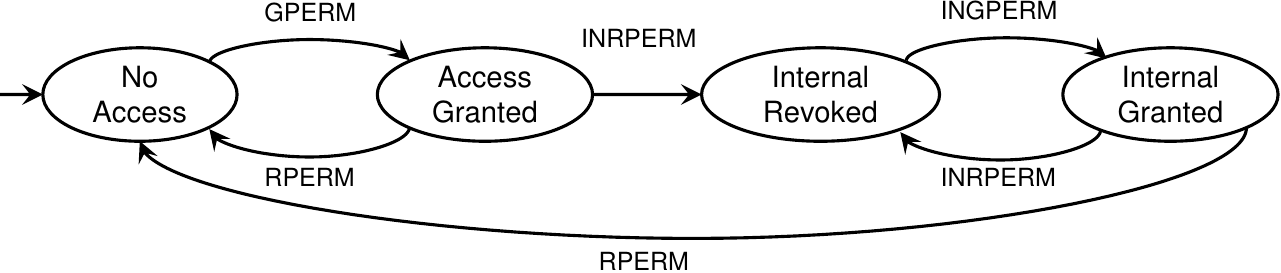}
  \caption{State machine for call gates.}
  \label{fig:state}
\end{figure}

\noindent \textbf{Call Gate Hardening.}
Attackers can potentially hijack non-crypto functions to jump to any position within crypto functions, including jumping directly to \texttt{INGPERM}. To secure that \texttt{INGPERM} is invoked through legitimate control flow, the \emph{\sysname-Monitor} enforces state transitions of call gates as per a defined state machine, depicted in \autoref{fig:state}. The execution of \texttt{INGPERM} is valid only after passing sequentially through the \emph{Access Granted} and \emph{Internal Revoked} states, ensuring control flow integrity of crypto functions.
Moreover, to secure that executing \texttt{INGPERM} to re-enter the crypto function corresponds strictly to its paired \texttt{INRPERM}, the \emph{\sysname-Monitor} verifies the address of \texttt{INGPERM} before executing it.

\noindent \textbf{Key-Related System Call Replacement.}
Since the OS is untrusted, system calls within crypto functions must be replaced with secure implementations. For memory management, while the OS remains responsible for allocating and mapping the crypto buffer, this memory is delegated to the \emph{\sysname-Monitor}, which manages the available memory within the buffer. As a result, the \emph{\sysname-Compiler} replaces the native \texttt{malloc} and \texttt{free} with secure versions implemented by the \emph{\sysname-Monitor}. Similarly, for all I/O operations involving keys and random number generation, the \emph{\sysname-Monitor} provides secure implementations, as detailed in \cref{subsec:io_key}.

\noindent \textbf{Iago Attack Defense.}
The OS can achieve Iago attacks~\cite{checkoway2013iago}, which could deceive the crypto function by manipulating the return values of system calls. Since non-sensitive memory allocation still uses native system calls (e.g., \texttt{mmap}), there's a risk that the returned memory area might overlap with the crypto buffer. To mitigate this, \sysname performs two key actions: it checks page table updates to prevent double mapping of the crypto buffer (\cref{subsec:exception}) and inserts runtime checks to ensure the returned memory does not overlap with the crypto buffer. Furthermore, system calls that return length data, such as \texttt{recv(socket, buf, len, flags)}, are verified to prevent exceeding the buffer's maximum length.

\begin{figure*}[!t]
  \centering
  \includegraphics[width=6.5 in]{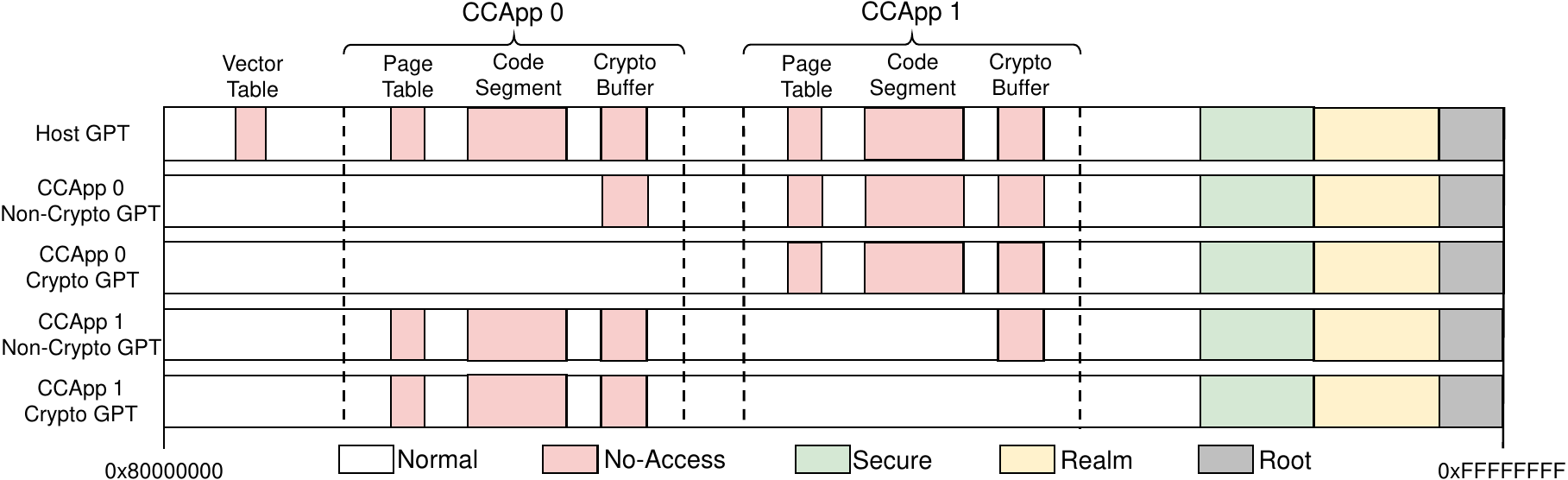}
  \caption{Physical memory access control views for privileged software, crypto functions, and non-crypto functions.}
  \label{fig:gpts}
\end{figure*}

\section{\sysname Runtime Isolation} \label{sec:isolation}

\subsection{Memory Isolation} \label{subsec:memory}

To minimize the TCB, memory allocation is managed by the untrusted \emph{\sysname-driver}. The \emph{\sysname-Monitor} then delegates this memory to \appnames by updating the GPT. However, GPT-based memory isolation faces three critical challenges:

\noindent \textbf{Multi-Core Isolation.}
CCA typically allows multiple cores to share the same GPT, maintaining a unified memory view. However, this approach can lead to permission conflicts: crypto functions on one core require different access permissions for the crypto buffer compared to the OS or non-crypto logic running on other cores. To address this, \sysname introduces three distinct types of GPTs, as depicted in \autoref{fig:gpts}. The Host GPT is designated for untrusted privileged software like the OS. Each \appname is assigned a Crypto GPT for handling keys within crypto functions, and a Non-Crypto GPT for non-crypto functions. By configuring each core with its own GPT, \sysname ensures that each core has a unique view of physical memory access permissions. The \emph{\sysname-Monitor} is responsible for initializing these GPTs and dynamically switching between them based on the execution context.

\noindent \textbf{Multi-GPT Maintenance.}
The multi-GPT mechanism disrupts CCA's native unified permission management, and the coexistence of these tables introduces additional requirements for consistency maintenance and security constraints. First, accessing these GPTs requires acquiring a spinlock to ensure the GPTs' views are not outdated. We use the Host GPT as a template, as it is configured to deny access to the isolated memory of all \appnames, requiring only minimal adjustments to create the GPTs for each \appname. To ensure that the delegated memory does not overlap—whether delegated to \appnames or to the original TEEs (including the secure world and realm world)—we stipulate that only \emph{normal} memory in the Host GPT can be delegated. Additionally, to enforce isolation between \appnames, the \emph{\sysname-Monitor}, when delegating memory to a \appname, configures that memory as \emph{non-access} in the GPTs of all other \appnames. Similarly, to prevent \appnames from accessing memory in the original TEEs, the \emph{\sysname-Monitor} sets the memory as \emph{realm} or \emph{secure} in the GPTs of all \appnames when memory is delegated to these TEEs.

\noindent \textbf{Mitigating the Semantic Gap.}
Since GPTs control access to physical memory while \appnames operate with a virtual address view, the OS might tamper with page tables to map sensitive areas like the crypto buffer to unprotected memory regions. To bridge this gap, after the OS creates the page tables, the \emph{\sysname-Monitor} first sets these tables to \emph{no-access} in the Host GPT and then scrutinizes the mappings in these tables; this order removes any TOCTTOU window~\cite{bratus2008toctou}.
It performs three critical checks: 
(1) \emph{Single Mapping Verification}: It ensures that the code segment and crypto buffer are correctly and uniquely mapped to the delegated physical memory. This setup prevents code-reuse attacks caused by page-oriented programming~\cite{hanpage2024pop} and protects against Iago attacks~\cite{checkoway2013iago}.
(2) \emph{Table Protection}: It confirms that the page table and vector table are protected from being maliciously mapped, defending against tampering by the untrusted logic of \appnames. 
(3) \emph{Executable Space Control}: The code segment is verified to be read-only and executable, while ensuring no other memory areas are executable, preventing code injection attacks. 
Additionally, if a page fault occurs during \appname execution, \sysname employs a shadow page table mechanism to enable the OS to securely update the \appname page table, as further detailed in \cref{subsec:exception}.

\subsection{Secure Context Switch} \label{subsec:exception}

In this section, we detail how \sysname handles call gate execution to achieve intra-process isolation and how it intercepts exceptions to effectively isolate \appnames from the OS.

\noindent \textbf{Call Gate Handling.}
The \emph{\sysname-Monitor} identifies the call gate executor using the GPT base address. Since the OS and untrusted user-space applications utilize the Host GPT, their attempts to invoke call gates are effectively blocked. We discuss the handling of call gates from three aspects:

(1) \emph{Call Gate Construction:} Call gates are initiated with a Supervisor Call (\texttt{svc}) that traps to EL1, followed by a Secure Monitor Call (\texttt{smc}) in the vector table that traps to EL3. Typically, the \texttt{svc} calling convention uses an immediate number set to 0, with the call number passed through the \texttt{x8} register. To prevent tampering with the register to manipulate call gates, our call gates use a different convention where the call number is passed directly through an immediate number (e.g., \texttt{svc \#4}).  Additionally, the \emph{\sysname-Compiler} ensures that all original \texttt{svc} immediate values in the \appname, before instrumentation, are set to 0 to prevent conflicts.

(2) \emph{Call Gate Routing:} The call gate routing is implemented through the previously delegated vector table. Upon trapping to EL3 via \texttt{svc} and \texttt{smc}, the \emph{\sysname-Monitor} switches to the relevant Crypto GPT or Non-Crypto GPT based on the current context of the \appname. After making this switch, it bypasses the OS and directly returns control to the \appname.

(3) \emph{Register State Sanitization:} 
This process includes entry status flag sanitization and exit register leakage prevention.
At the entry of crypto functions,
the \emph{\sysname-Monitor} configures the \texttt{PSTATE} register by clearing all condition flags and ensuring key status bits are set as \emph{invariants}, such as ensuring the debugging bit is disabled.
At the exit of crypto functions, the \emph{\sysname-Monitor} cleans the general register states to prevent any inadvertent leakage of sensitive data.

\noindent \textbf{Regular Exception Interception.}
\sysname also intercepts regular exceptions through the delegated vector table, ensuring that any execution flow leaving the \appname first enters the \emph{\sysname-Monitor} for isolation from the OS. We execute three crucial operations to enforce this isolation:

(1) \emph{GPT Switching}: To prevent OS access to isolated memory, the \emph{\sysname-Monitor} switches to the Host GPT prior to entering the OS and switches back to the relevant GPT upon entering the \appname. 
Note that even if the OS directly jumps back to the \appname, this will trigger a GPF due to lacking permissions to the code and crypto buffer of the \appname.

(2) \emph{Sensitive Context Cleanup}: To protect against the OS spying on register states via exceptions or modifying the exception return address to hijack control flow, the \emph{\sysname-Monitor} preserves sensitive context of the crypto function. This preservation includes general registers, the exception link register, and the saved program status register. Before handing control to the OS, it clears the general registers (except for the parameter registers, frame pointer, and link registers) and later restores the saved sensitive context upon re-entering the \appname. 
Furthermore, to ensure that exceptions during \appname execution are consistently intercepted, the delegated vector table is reactivated each time the \appname is re-entered.

(3) \emph{\appname Page Table Updating}: 
To allow normal updates to the \appname page tables, the OS prepares a shadow page table before delegating the \appname page table. When handling a page fault, the \emph{\sysname-Monitor} temporarily switches to this shadow page table, permitting the OS to update it normally. Before resuming \appname execution, the \emph{\sysname-Monitor} reverts to the delegated \appname page tables and synchronizes changes made to the shadow page table. 
It also conducts checks on new mappings, similar to those performed during initial \appname page table delegation. Importantly, the \emph{\sysname-Monitor} ensures that newly mapped memory does not overlap with the crypto buffer, particularly the crypto stack. This verification ensures that the runtime checks (\cref{subsec:instrumentation}) for system call return values, effectively defend against memory-based Iago attacks~\cite{checkoway2013iago}. Further defenses against other Iago attacks can incorporate established methods~\cite{shinde2020besfs}.

\subsection{Sensitive I/O and Key Generation} \label{subsec:io_key}

\begin{figure}[!ht]
  \centering
  \includegraphics[width=3.2 in]{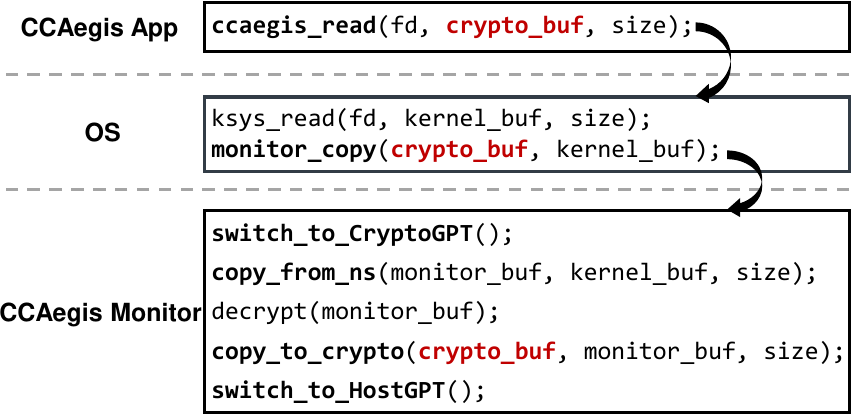}
  \caption{Encapsulated key loading of \sysname.}
  \label{fig:io}
\end{figure}

Since the OS is untrusted, the \emph{\sysname-Monitor} implements protections for key generation, import, and export within \appnames. For key generation, the process follows the call gate routing, which bypasses the OS and is managed directly by the \emph{\sysname-Monitor}. By utilizing the Random Number Generator (RNG) feature in Armv8.5-A, the \emph{\sysname-Monitor} generates true random numbers by reading the \texttt{RNDR} register and writing its value into the crypto buffer. Key import from the file system, as depicted in \autoref{fig:io}, begins with the \appname requesting the OS to read encapsulated keys from a file. Since the OS operates under the Host GPT, it must request the \emph{\sysname-Monitor} to decrypt the keys and transfer them into the crypto buffer. To safeguard against unauthorized access: (1) Before decrypting the keys, the \emph{\sysname-Monitor} verifies the file signature to ensure it belongs to the current \appname, thus preventing the \texttt{monitor\_copy} interface from being exploited as a decryption oracle. (2) The \emph{\sysname-Monitor} implements the \texttt{copy\_from\_ns} to verify that the kernel buffer being copied is classified as \emph{normal} memory in the Host GPT, preventing the OS from using this interface to steal secrets from other worlds or the crypto buffer. (3) It also implements the \texttt{copy\_to\_crypto} to prevent key leakage to unprotected memory. The process for exporting keys follows a similar procedure and is therefore not detailed here.

\subsection{Multi-threading Support.} \label{subsec:multi-thread}

\sysname supports thread synchronization primitives (such as locks, conditional variables, and semaphores) based on Fast User-space muTEX (Futex). The OS can access the Futex in the \appname memory normally since this variable is located outside the crypto buffer. Moreover, before executing external calls such as \texttt{FUTEX\_WAIT}, \sysname executes \texttt{INRPERM} to prevent potential key leakage attacks (\cref{subsec:instrumentation}). 
For each new \appname thread, a separate crypto stack is allocated. The \emph{\sysname-Monitor}  checks the mappings of these stacks, similar to checks in the page table delegation (\cref{subsec:memory}). During exception handling, the \emph{\sysname-Monitor} saves the sensitive state for each thread and restores the appropriate sensitive state based on the thread ID before re-entering the \appname.
To support concurrency efficiently, each core maintains its own GPT view so a domain switch is per core: the \emph{\sysname-Monitor} writes that core’s GPT base, performs a local TLB invalidation, and returns to user mode. This hot path is lock free across cores, and we disable TLB sharing (clear CnP) to avoid cross-core shootdowns during switches. A global spinlock is taken only when GPT structures are updated or memory is delegated, which occurs infrequently, and multi-GPT maintenance templates new GPTs from the Host GPT to keep update time short (\cref{subsec:memory}).

\section{Implementation} \label{sec:prototype}

Current hardware lacks Arm CCA support, prompting development of both functional and performance prototypes.

\noindent \textbf{Functional Prototype.} 
We implemented a prototype on the Arm Fixed Virtual Platform (FVP)~\cite{arm2023fvp}, which supports RME and RNG features, to validate the design of \sysname and confirm its compatibility with future hardware.
FVP has been used in previous works~\cite{zhang2023shelter,sridhara2024acai,wang2024cage} to test the functional correctness of CCA. 
The static analysis module of \sysname utilizes LLVM to analyze and transform the source code of \appnames, integrating a taint analyzer and an LLVM Pass, adding 2k LoCs. Our analysis preserve a fail-closed default: when precision is insufficient at a boundary, we conservatively mark it sensitive. If a call target on a tainted path cannot be resolved statically, including calls reached via \texttt{dlopen}/\texttt{dlsym}, we treat that site as a sensitive boundary, auto-insert domain switches, and place outputs and reachable buffers in the protected region; this conservative choice may enlarge the sensitive slice, which we mitigate with function-boundary hoisting and batching (\cref{subsec:instrumentation}).
Additionally, \sysname modifies the Linux kernel 6.6-rc6 and introduces a Linux driver, adding another 1k LoCs. This driver manages the lifecycle of \appnames, interacting with the \emph{\sysname-Monitor} via \texttt{smc} to handle the creation and destruction of \appnames, and the creation of new threads. Furthermore, \sysname develops a loader that utilizes the \texttt{ioctl} interface provided by the driver to load instrumented \appnames. The driver is also responsible for \appname memory allocation and mapping, utilizing the Contiguous Memory Allocator (CMA) to manage three distinct contiguous physical memory areas assigned for the code segment, page table, and crypto buffer. Upon driver initiation, the Linux kernel completes the vector table registration, ensuring that any control flow exiting the \appname is initially intercepted. In the root world, the \emph{\sysname-Monitor} exposes a narrow SMC ABI that configures/switches GPT domains and verifies shadow mappings (no scheduler, filesystem, or device stacks). This fail-closed design keeps the trusted code path compact and auditable, and in practice amounts to a small, targeted firmware extension, not a rewrite.

\noindent \textbf{Performance Prototype.}
As the FVP is not cycle-accurate, we evaluate performance using the Rock Pi 4B development board (Armv8.0-A). The cores used in this development board are consistent with those in previous works~\cite{zhang2023shelter,wang2024cage,sridhara2024acai}. However, since the hardware features (i.e., GPC and RNG) used in this paper are not supported on this development board, we simulate these features as follows:
(1) Following previous work~\cite{sridhara2024acai}, we utilize idle registers in EL3—specifically, \texttt{afsr0\_el3} and \texttt{afsr1\_el3}—to substitute for the \texttt{gpccr\_el3} (the GPC control register) and \texttt{gptbr\_el3} (the GPT base register) for evaluating the overhead of GPC configuration and GPT base address switching.
Since the GPT is an in-memory structure, we consistently implement and assess the overhead of creating, maintaining, and destroying the GPTs like the functional prototype.
(2) We replace the TLB maintenance instruction (\texttt{TLBI PAALLOS}, not available on our board), which typically only invalidates all cached GPT information in the TLB, with an instruction that invalidates the entire TLB cache. 
(3) We generate a seemingly random 64-bit key using the stack pointer (\texttt{sp}), the link register (\texttt{x30}), and the counter-timer physical count register (\texttt{cntpct\_el0}) to simulate the RNG feature.

\section{Evaluation} \label{sec:security_evaluation}
To evaluate the security and performance of \sysname, we aim to answer the following five questions:

\noindent \textbf{Q1:} How large is the TCB of \sysname? 

\noindent \textbf{Q2:} Can \sysname defeat privileged and intra-process attacks?

\noindent \textbf{Q3:} What is the compilation overhead of our static analysis?

\noindent \textbf{Q4:} What is the performance cost of our fine-grained isolation measured on micro-benchmarks and real-world applications?

\noindent \textbf{Q5:} What is the root cause of the performance overhead?

\subsection{Experimental Setup} \label{subsec:setup}
We evaluate the security of \sysname on the functional prototype implemented on FVP, answering \textbf{Q1} and \textbf{Q2}. We use LLVM 10.0.1 with -O2 optimizations to implement static analysis and evaluate the compile-time overhead to answer \textbf{Q3}. It operates on an Arm server powered by a 96-core HiSilicon Kunpeng-920 CPU (2.6 GHz, Armv8.2-A) and is equipped with 256 GB RAM. We assess the performance overhead of \sysname on the performance prototype implemented on the Rock Pi 4B development board, answering \textbf{Q4} and \textbf{Q5}. This board is built around a Rockchip RK3399 SoC, featuring two Arm Cortex-A72 cores (running at speeds of up to 1.8GHz) and four Cortex-A53 cores (up to 1.4GHz). To avoid performance measurement discrepancies caused by the big.LITTLE architecture of this SoC, we set the Cortex-A72 cores to run at the highest frequency and disabled all Cortex-A53 cores.

\subsection{Q1: TCB Size of \sysname} \label{subsec:tcb}

\begin{table}[!htbp]
    \caption{TCB modifications of \sysname.}
  \label{tab:LOC}
  \centering
  \begin{tabular}{lc}
  \hline

  \hline
  \textbf{Component} & \textbf{Lines of Code (LoCs)} \\
  \hline
    Memory Isolation & 990\\

    Exception Handling Interception & 483\\

    Secure I/O and Key Generation & 154\\

    Other Configuration & 176 \\
  \hline
    \textbf{All} & \textbf{1803} \\
  \hline

  \hline
  \end{tabular}
\end{table}

We run the \texttt{cloc} tool to measure the code size of \sysname. The TCB of \sysname is confined to the Monitor located in the root world, which uses Trusted Firmware-A v2.9.0~\cite{arm2024tfa}, accounting for 479K LoCs. \sysname introduces an additional 1,803 LoCs of modifications, as detailed in \autoref{tab:LOC}. 
Unlike CCA's heavier TCB, \sysname omits Linux cca-guest-v2~\cite{arm2024ccaguest} within CVMs, with its 26M LoCs, and TF-RMM v0.4.0~\cite{arm2023tfrmm}, which includes 33K LoCs.

\subsection{Q2: Security of \sysname} \label{subsec:security}

\begin{table}[!htbp]
	\centering
    \caption{Main attacks and defenses of \sysname.}
    \label{tab:attacks}
        \begin{adjustbox}{max width=\linewidth}
	\begin{tabular}{ll}
        \hline

        \hline
	   \textbf{Attack} & \textbf{\sysname defence}   \\
        \hline

        \multicolumn{2}{l}{\textbf{From privileged software:}} \\
          ~\textbf{At creation-time} \\
            ~\ding{182} Incorrect binary loading & Monitor checks  \\
            ~\ding{183} Overlapped memory delegation & GPT access synchronization \\
          ~\textbf{At runtime} \\
            ~\ding{184} CPU GPC circumvention  & EL3 isolation and TLB invalidation \\
            ~\ding{185} Unauthorized memory access & Memory delegation and GPC \\
            ~\ding{186} Malicious I/O and entropy & System call replacement \\
            ~\ding{187} Exception interception circumvention & Vector table isolation and GPC \\
            ~\ding{188} Register leakage & Exception context sanitization \\
            ~\ding{189} Illegal mappings and Iago & Monitor checks and runtime checks \\
            ~\ding{190} Call gates invocation & Monitor checks \\
          ~\textbf{At destruction-time} \\
            ~\ding{191} Sensitive data remanence & Monitor cleanup \\
        \hline
        \multicolumn{2}{l}{\textbf{From non-crypto functions:}} \\
            ~\ding{182} Unauthorized memory access & Page table checks and GPC \\
            ~\ding{183} Control flow hijacking & Monitor checks and GPC \\
            ~\ding{184} Register leakage & Call gate context sanitization \\
        \hline
        \multicolumn{2}{l}{\textbf{From malicious \appname:}} \\
            ~\ding{182} Collusion with OS & Monitor checks and GPC \\
        \hline
        \multicolumn{2}{l}{\textbf{From peripherals:}} \\
            ~\ding{182} Malicious DMA & SMMU GPC \\
            ~\ding{183} Peripheral GPC circumvention & EL3 isolation and TLB invalidation \\
        \hline

        \hline
	\end{tabular}
        \end{adjustbox}
	\label{tab:defense}
\end{table}

\begin{table*}[!t]
    \caption{The number and time of annotations, TCB size, call gates, syscall replacements, and the compile time for \sysname.}
  \label{tab:analysis}
  \centering
  \begin{adjustbox}{max width=\linewidth}
\begin{tabular}{lccccccccccccc}
  \hline

  \hline
\multirow{2}{*}{\textbf{Application}} & \textbf{Number of Tag} & \textbf{Annotation} & \textbf{Total} & \textbf{Crypto} & \textbf{Total} & \textbf{Crypto} & \textbf{False} & \textbf{Call} & \textbf{Internal} & \textbf{Syscall} & \textbf{Original} & \textbf{\sysname}\\
 & \textbf{(Source + Sink)} & \textbf{Time (h)} & \textbf{LoC} & \textbf{LoC} & \textbf{Function} & \textbf{Function} & \textbf{Positive} & \textbf{Gate} & \textbf{Call Gate} & \textbf{Replace} & \textbf{Time (s)} & \textbf{Time (s)}\\
 \hline
  \textbf{\texttt{wolfSSL}} & 8 + 5 & 5.48 & 174067 & 41384 (23.77\%) & 2700 & 343 (12.70\%) & 46 (1.91\%) & 704 & 2534 & 3 & 34.41 & 103.06 & \\
  \textbf{\texttt{ccrypt}} & 6 + 3 & 2.25 & 5670 & 1156 (20.38\%) & 39 & 21 (53.85\%) & 3 (14.29\%) & 295 & 126 & 12 & 2.52 & 15.84 & \\
  \textbf{\texttt{libhydrogen}} & 2 + 4 & 2.85 & 2995 & 1056 (35.26\%) & 75 & 40 (53.33\%) & 3 (7.89\%) & 108 & 236 & 2 & 2.33 & 4.43 & \\
  \textbf{\texttt{libxcrypt}} & 3 + 2 & 1.72 & 16649 & 1759 (10.57\%) & 121 & 8 (6.61\%) & 0 (0.00\%) & 172 & 180 & 3 & 12.34 & 21.95 & \\
  \hline

  \hline
\end{tabular}
\end{adjustbox}
\end{table*}

As shown in \autoref{tab:attacks}, we categorize the security risks and demonstrate how our defense measures address them.

\noindent \textbf{Privileged Software.}
\sysname protects \appname initialization from the untrusted OS. Specifically, \ding{182} after the untrusted OS loads the \appname binary, the \emph{\sysname-Monitor} verifies the loading address and the integrity of the binary. 
\ding{183} Before allocating memory to \appnames or the original TEEs, the \emph{\sysname-Monitor} ensures the memory is allocable by consulting the Host GPT. To avoid memory overlaps due to outdated views, it synchronizes access to all GPTs across cores (\cref{subsec:memory}).

We further outline the corresponding attacks and countermeasures during the execution of \appname. \ding{184} Since the GPT is stored in the root world and the GPC registers reside at the EL3 level, privileged software cannot disable the GPC or modify the GPT. Given that the GPT may be cached in the TLB, the \emph{\sysname-Monitor} invalidates the TLB during GPT updates or when switching GPT base addresses. As the TLB may be shared across multiple cores, the \emph{\sysname-Monitor} clears the \texttt{CnP} bit in the \texttt{TTBR} registers to prevent TLB sharing.
\ding{185} Memory areas like the code, crypto buffer, and page tables of the \appname are isolated with GPTs such that unauthorized access by privileged software will trigger a GPF. 
\ding{186} All random number generation and key-related I/O are secured by implementations in the \emph{\sysname-Monitor}, and direct reads of encapsulated key files do not result in leakage (\cref{subsec:io_key}).
\ding{187} During \appname execution, the \emph{\sysname-Monitor} intercepts exceptions via a delegated vector table, managing GPT switching to ensure proper isolation (\cref{subsec:exception}). Modifications to this vector table will also trigger a GPF. To guarantee that this interception mechanism is not circumvented, the \emph{\sysname-Monitor} activates the vector table each time the \appname is re-entered.
\ding{188} The \emph{\sysname-Monitor} also sanitizes the sensitive register states of crypto functions during interceptions to prevent key leakage. Additionally, it manages exception returns to prevent tampering with return addresses.
\ding{189} Furthermore, the \emph{\sysname-Monitor} reviews page table updates to block malicious mappings and the \emph{\sysname-Compiler} integrates runtime checks of system call return values to counter Iago attacks~\cite{checkoway2013iago}.  
\ding{190} To prevent privileged software from calling the interfaces provided by the \emph{\sysname-Monitor}, such as calling the call gate to grant access to the crypto buffer, the \emph{\sysname-Monitor} checks \texttt{gptbr\_el3}.

\sysname also protects the destruction phase. Specifically, \ding{191} upon destroying a \appname, \sysname clears its crypto buffer, cache, and register states, then releases the delegated physical memory areas (i.e., set as \emph{normal}) from the GPTs of the host and other \appnames, and finally destroys the \appname's GPTs.
 
\noindent \textbf{Non-Crypto Functions.}
\ding{182} The \emph{\sysname-compiler} inserts call gates into crypto functions to ensure that any access to the crypto buffer by non-crypto functions triggers a GPF. As depicted in \autoref{fig:gpts}, the vector table, page table, and \appname code are not marked as \emph{no-access} in the Non-Crypto GPT. Instead, the \emph{\sysname-Monitor} safeguards these components by checking updates to the page table (\cref{subsec:exception}).
\ding{183} To prevent attackers from gaining access to the crypto buffer through control flow hijacking, the \emph{\sysname-compiler} inserts \texttt{INRPERM} before all untrusted calls to revoke access permissions. Moreover, the \emph{\sysname-Monitor} manages the state transitions of call gates to ensure that call gates are invoked through legitimate control flow (\cref{subsec:instrumentation}).
\ding{184} The \emph{\sysname-Monitor} clears the registers before exiting the crypto function or invoking untrusted calls to prevent key leakage.

\noindent \textbf{Malicious \appname.}
\ding{182} An \appname can potentially collude with the OS to allocate memory from other \appnames or worlds to itself. The \emph{\sysname-Monitor} verifies the memory to be delegated, ensuring no overlap occurs.

\noindent \textbf{Peripherals.}
\ding{182} Attackers can use DMA to bypass the CPU GPC and directly access isolated memory. The \emph{\sysname-Monitor} utilizes the SMMU to enable GPC for all peripherals. It configures the Host GPT for SMMU and ensures that peripherals maintain the lowest access permissions.
\ding{183} The SMMU GPC can only be configured by the root world. When modifying the Host GPT, the \emph{\sysname-Monitor} also invalidates the SMMU TLB, preventing peripheral GPC bypass.

\subsection{Q3: Evaluation on Static Analysis} \label{subsec:analysis_overhead}

We considered four widely used cryptographic libraries (i.e., \texttt{wolfSSL}, \texttt{ccrypt}, \texttt{libhydrogen}, \texttt{libxcrypt}) to test the static analysis. 
For each library, we evaluated the number of manual annotations as well as the results of automated program partitioning and instrumentation.

\noindent \textbf{Annotated Tags.} 
We designate the keys in cryptographic algorithms as \emph{taint sources} and the outputs of these algorithms (including encrypted ciphertext, decrypted plaintext, and signatures) as \emph{taint sinks}. The LoCs and time required for annotations in the cryptographic libraries are described in \autoref{tab:analysis}. 
We invited five graduate students to understand the semantics of API parameters and annotate them. For most libraries, the task took on average less than three hours (experienced developers could likely complete it faster). For larger cryptographic libraries with extensive APIs, such as \texttt{wolfSSL}, the annotation process required under six hours.
This level of effort is acceptable, which indicates that \sysname has good practicality and deployment friendliness. For \texttt{wolfSSL}, we added annotations in 13 LoCs, which include \textbf{eight} \emph{taint sources} on keys used in the \texttt{AesGcm} and \texttt{AesCfb} APIs, along with their associated \texttt{Aes} context structures. Additionally, we annotated \textbf{five} \emph{taint sinks} on ciphertext variables associated with these contexts. In \texttt{ccrypt}, we applied annotations in \textbf{six} LoCs for \emph{taint sources} on command-line key inputs (\texttt{cmd.keyword}) and internal cryptographic states, such as \texttt{block2} in \texttt{xrijndaelEncrypt}, plus \textbf{three} \emph{taint sinks} for the stream inputs of \texttt{streamhandler} functions that process encrypted data. For \texttt{libhydrogen}, the annotations were minimal, involving only \textbf{two} \emph{taint sources} on secret keys in the \texttt{hydro\_sign\_create} and \texttt{hydro\_secretbox\_setup}, and \textbf{four} \emph{taint sinks} on their output buffers, signatures, and internal states. In \texttt{libxcrypt}, we used \textbf{three} \emph{taint sources} on \texttt{HMAC} keys and \textbf{two} \emph{taint sinks} on the outputs of \texttt{hmac\_sha1\_process\_data} functions.

\noindent \textbf{False Positive Analysis.}
As noted in prior studies~\cite{hardekopf2011flow,gharat2016flow}, points-to analysis in C programs is inherently imprecise due to its undecidability~\cite{ramalingam1994undecidability}. Consequently, some non-crypto functions may inadvertently be included within the crypto function set, resulting in false positives. While this leads to slight over-protection, it ensures no genuine crypto functions are missed, thereby preventing key leakage. As shown in \autoref{tab:analysis}, for \texttt{wolfSSL}, 343 functions (12.70\%) were identified as crypto functions, constituting only 23.77\% of the total LoCs. Similar trends are observed in other cryptographic libraries, where crypto code ranges between 10.57\% and 35.26\% of the total LoCs. In the absence of ground truth, we manually verified false-positive instances. Our evaluation revealed zero false positives for \texttt{libxcrypt}, and a low false-positive rate of 1.91\% for \texttt{wolfSSL}. Both \texttt{ccrypt} and \texttt{libhydrogen} exhibited higher false-positive rates (14.29\% and 7.89\%, respectively), primarily due to their small total number of crypto functions—indeed, only three functions across these libraries were incorrectly flagged as crypto functions.

\noindent \textbf{Compilation Overhead.} 
We evaluated these workloads by focusing on three aspects: the number of call gates inserted, the number of system call replacements, and the overall compilation time. Regarding call gate insertion, \texttt{wolfSSL} exhibited the highest number, as shown in \autoref{tab:analysis}, with a total of 704 call gates and 2,534 internal call gates inserted. The number of domain-switching instructions directly impacts execution efficiency; this relationship is further analyzed in \cref{subsec:root_cause}. Additionally, key-related system calls were replaced to ensure secure handling; for instance, \texttt{ccrypt} required 12 system call replacements (primarily \texttt{malloc} and \texttt{free}), as it mainly utilizes the crypto heap for key storage. The compilation time overhead introduced is moderate, with the largest overhead observed in \texttt{wolfSSL}, which adds approximately one minute.

\subsection{Q4: Evaluation on Benchmarks and Applications} \label{subsec:microbenchmarks}

\noindent \textbf{Creation and Destruction Cost.}
We measured the basic costs incurred during the creation and destruction phases of an \appname by repeatedly launching and terminating an empty application 10,000 times. On average, when the driver is loaded, \sysname delegates the vector table, resulting in an overhead of about 1 ms. During the \appname loading phase, \sysname creates Crypto and Non-Crypto GPTs, as well as verifies page tables and program code (including dynamic libraries), adding an overhead of 16.99 ms. Correspondingly, during the destruction phase, \sysname releases the delegated memory and destroys the established GPTs, incurring an additional overhead of 1.90 ms. Nevertheless, these are all one-time costs that do not affect the runtime performance. The primary runtime overhead arises from the following domain-switching operations.

\noindent \textbf{Domain Switching Overhead.}
We measured the overhead of domain switching in CPU cycles using the Performance Monitoring Unit (PMU) over one million iterations. The average overhead was 1,211 cycles for \texttt{GPERM} and 1,269 cycles for \texttt{RPERM}. This overhead primarily arises from expensive TLB flushes associated with GPT switches and the context save-and-restore triggered by traps to EL3. Internal call gates within crypto functions exhibited similar overheads. Additionally, we measured the extra overhead introduced by traps when exiting and entering the \appname for exception handling (e.g., system calls), which were 1,067 and 1,437 cycles, respectively.

\begin{figure}[!htbp]
  \centering
  \includegraphics[width=3.3 in]{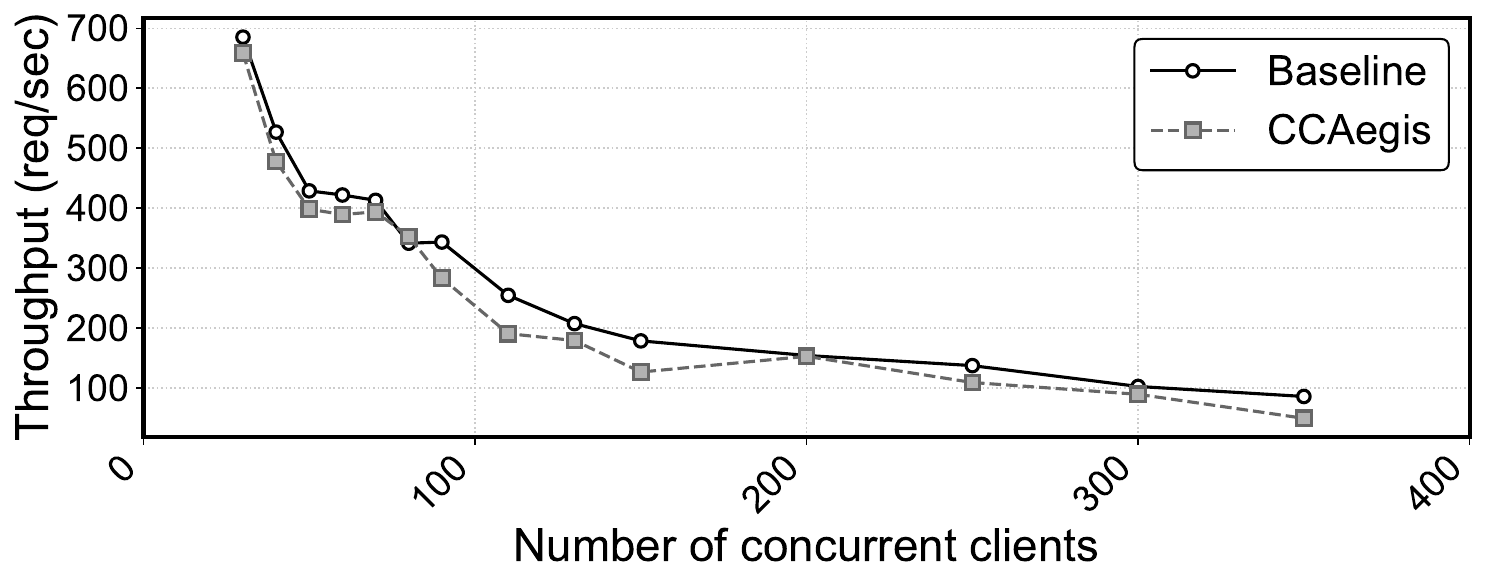}
  \caption{Throughput cost on \texttt{Nginx} incurred by \sysname.}
  \label{fig:concurrent}
\end{figure}

\noindent \textbf{\texttt{Nginx} Throughput.}
Following previous studies~\cite{jin2022annotating,zhang2023shelter,liu2015thwarting}, we used the \texttt{Nginx v1.21.4} web server to measure the impact of \sysname's protection on throughput. As our baseline, we employed a \texttt{Nginx} server compiled with an uninstrumented \texttt{wolfSSL v5.1.0} using the \texttt{ECDHE\_RSA\_WITH\_AES\_128\_GCM\_SHA256} cipher suite, without using the \sysname loader. We conducted the evaluation using two machines connected via LAN: one configured as the \texttt{Nginx} server, and the other running \texttt{Apache Bench (ab) v2.3} to simulate a client sending 10,000 requests with a file size of 32KB. Since the overhead primarily arises during the connection establishment phase, we enabled the \texttt{keepalive} option to more accurately measure overhead during request processing, aligning closely with realistic usage scenarios. To compare the throughput overhead under high and low concurrency, we incrementally increased the number of concurrent clients up to the system's maximum capacity (350 concurrent clients, beyond which significant connection failures occurred). As illustrated in \autoref{fig:concurrent}, under low concurrency conditions ($\le$80 clients), the overhead introduced by \sysname remained below 10\%, peaking at 9.33\% with 40 concurrent clients. However, as concurrency increased further, the throughput overhead rose sharply, reaching a maximum of 42.72\% at 350 concurrent clients.

\noindent \textbf{\texttt{wolfSSH} Latency.}
To evaluate the overhead \sysname introduces to SSH connections, we used a \texttt{wolfSSH v1.4.17} client (compiled with the instrumented \texttt{wolfSSL} cryptographic library) to establish 1,000 connections to a local \texttt{OpenSSH v8.2p1} server, comparing the results against an uninstrumented baseline. On average, \sysname’s protection incurred a 25.35\% overhead per connection request, corresponding to approximately 27.2 ms. Note that this overhead applies primarily during the initial connection phase; subsequent data transfers after the connection is established would exhibit significantly lower latency.

\begin{table}[!htbp]
    \caption{Cost of \sysname in \texttt{ccrypt}, \texttt{libhydrogen}, and \texttt{libxcrypt}.}
  \label{tab:libraries}
  \centering
    \begin{adjustbox}{max width=\linewidth}
  \begin{tabular}{cccccc}
  \hline

  \hline
  & \textbf{ccrypt} & \textbf{ccrypt} & \textbf{libhydrogen} & \textbf{libhydrogen} & \textbf{libxcrypt} \\
  & \textbf{encrypt} & \textbf{decrypt} & \textbf{encrypt} & \textbf{decrypt} & \textbf{SHA} \\
  \hline
  \textbf{Baseline (s)} & 0.5552 & 0.5587 & 0.4361 & 0.2168 & 0.0488 \\
  \textbf{\sysname (s)} & 0.5835 & 0.5830 & 0.4388 & 0.2175 & 0.0492\\
  \textbf{Overhead} & 5.10\% & 4.35\% & 0.61\% & 0.31\% & 0.84\% \\
  \hline

  \hline
  \end{tabular}
  \end{adjustbox}
\end{table}

\noindent \textbf{Cryptographic Libraries Performance.}
In addition to \texttt{wolfSSL}, we further evaluated the performance overhead on other libraries: \texttt{ccrypt v1.11}, \texttt{libhydrogen} (commit \texttt{7ff9582}), and \texttt{libxcrypt v4.4.36}. To measure overhead, We modified the libraries' built-in test suites. Specifically, \texttt{libhydrogen} employs the \texttt{Gimli} permutation for encryption, \texttt{ccrypt} utilizes the \texttt{Rijndael} cipher, and \texttt{libxcrypt} implements \texttt{HMAC-SHA1}. Each test was performed on 10MB of data, repeated 1,000 times, with results summarized in \autoref{tab:libraries}. Overall, \sysname introduced low overhead across all three cryptographic libraries, with negligible performance impact ($<$1\%) on \texttt{libhydrogen} and \texttt{libxcrypt}, and moderate overhead on \texttt{ccrypt}—5.10\% for encryption and 4.35\% for decryption.

\noindent \textbf{Comparison with the State-of-the-Art.}
We compared \sysname with state-of-the-art CCA-based systems, specifically \textsc{Shelter}~\cite{zhang2023shelter}. Since \textsc{Shelter} was developed on the Juno R2 development board and its performance prototype is not open-sourced, we modified the functional prototype and ported it to the Rock Pi 4B development board, which we currently use. We evaluated the performance overhead on all the real-world applications previously tested. Compared to \textsc{Shelter}'s process-level isolation, \sysname provides intra-process isolation. However, achieving this fine-grained isolation with \sysname introduces additional overhead due to the insertion of call gates for isolation domain switching. Relative to \textsc{Shelter}, \sysname imposes less than 0.83\% additional overhead for cryptographic libraries like \texttt{ccrypt}, 7.43\% for \texttt{wolfSSH}, and the highest overhead for \texttt{Nginx} at 16.92\%.

\subsection{Q5: Root Cause Analysis} \label{subsec:root_cause}

\begin{figure}[!htbp]
  \centering
  \includegraphics[width=3.2 in]{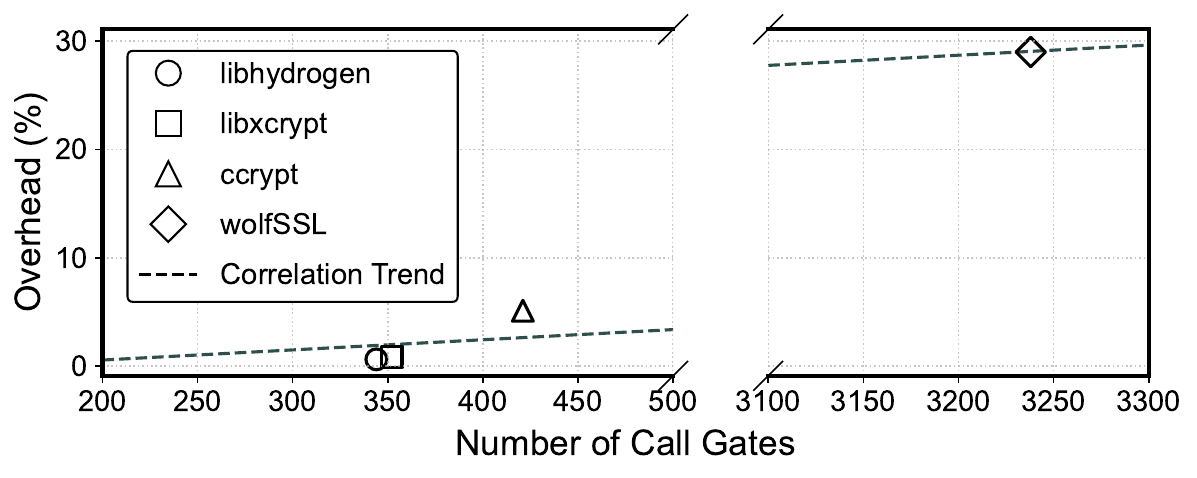}
  \caption{Correlation between number of call gates and overhead.}
  \label{fig:scatter}
\end{figure}

Through comparative experiments, we observed that \sysname's performance overhead is influenced by two key factors.
First, as shown in \autoref{fig:scatter}, the overhead increases with the number of call gates inserted to create isolated domains. For example, \texttt{wolfSSL}—with 3,238 call gates—experiences higher overhead (over 20\% greater) compared to other cryptographic libraries with fewer than 500 instrumented call gates. Second, the invocation frequency of call gates also directly affects execution efficiency. Even libraries with similar levels of instrumentation exhibit higher overhead under high-pressure workloads. As illustrated in \autoref{fig:concurrent}, high-concurrency scenarios trigger more frequent domain switches, leading to a threefold increase in overhead compared to low-concurrency conditions. We further conducted a stress test on \texttt{wolfSSL}'s AES-GCM. As depicted in \autoref{tab:aes}, this direct benchmarking of crypto functions revealed an even greater overhead (up to 75.80\%) compared to the overhead observed in high-concurrency \texttt{Nginx} scenarios invoking crypto functions.

\begin{table}[!t]
    \caption{Stress test of \texttt{wolfSSL}'s AES-GCM with a 2048-byte input size.}
  \label{tab:aes}
  \centering
    \begin{adjustbox}{max width=\linewidth}
  \begin{tabular}{ccccccc}
  \hline

  \hline
  & \textbf{AES128} & \textbf{AES128} & \textbf{AES192} & \textbf{AES192} & \textbf{AES256} & \textbf{AES256} \\
  & \textbf{encrypt} & \textbf{decrypt} & \textbf{encrypt} & \textbf{decrypt} & \textbf{encrypt} & \textbf{decrypt} \\
  \hline
  \textbf{Baseline (\textmu s)} & 36.36 & 36.36 & 39.76 & 39.72 & 43.31 & 43.27 \\
  \textbf{\sysname (\textmu s)} & 63.73 & 63.91 & 66.97 & 66.95 & 70.45 & 70.42 \\
  \textbf{Overhead} & 75.28\% & 75.80\% & 68.44\% & 68.56\% & 62.68\% & 62.74\% \\
  \hline

  \hline
  \end{tabular}
  \end{adjustbox}
\end{table}

\section{Discussion} \label{sec:discussion}

\noindent \textbf{Multi-Domain Support.}
\sysname currently supports only a single crypto domain, but it can be easily extended to support multiple independent domains. Separate GPTs can be created for function sets handling different types of secrets, ensuring that only the relevant functions have access to the memory.

\noindent \textbf{Performance Optimization.} Internal call gates secure untrusted calls within crypto functions, but frequent trapping to EL3 can be costly. To optimize efficiency without compromising security, we can perform security verification~\cite{haller2013dowsing} for frequently called external functions instead of instrumentation. Another practical optimization is to combine GPT enforcement with lightweight intra-process guards such as Pointer Authentication Code (PAC). GPT remains the authoritative barrier that keeps the OS and other processes from reading protected pages, while PAC provides a user-mode fast path that reduces \texttt{svc+smc} transitions inside a sensitive region.

\noindent \textbf{Scalability.}
\sysname primarily targets cryptographic applications because they stress-test intra-process isolation: secrets are repeatedly derived, transformed, and propagated across many small functions and library boundaries, which enlarges the taint surface and increases potential domain switches. By contrast, many noncryptographic scenarios (e.g., an ML service reading an API token) load a secret once and use it sparingly, where simple static labeling is often sufficient. Keys and their derivatives traverse multiple call chains and buffers, which is exactly where our taint guided function partitioning and GPT based enforcement are exercised most, so the reported overheads serve as a conservative upper bound for many general-purpose workloads. The design can be extended to other security-sensitive applications by adding monitor support for system calls beyond I/O and key generation (\cref{subsec:io_key}).

\noindent \textbf{False Negatives.}
While we report false positives in \cref{subsec:analysis_overhead}, we did not emphasize false negatives because our analysis prioritizes soundness over completeness. We adopt a fail closed default: when precision is insufficient at a boundary, we conservatively mark it sensitive. For ground truth, we manually validated the false positives, but quantifying false negatives is difficult because real applications lack an oracle that labels all true secret flows. We do not claim to eliminate all false negatives, and complete soundness for complex programs is undecidable. Potential sources include implicit or indirect control transfers that obscure flows, assembly with branches or jumps, and missing taint or erase annotations. As future work, we will ship default signatures for common cryptographic APIs and emit warnings on control flow redirections in assembly or indirect jumps.

\section{Related Work} \label{sec:related_work}

\noindent \textbf{Compartmentalization.}
Partitioning a program and isolating the secure parts is an effective way to enforce the principle of least privilege. To assist developers in program partitioning, SOAAP~\cite{gudka2015clean} provides an interactive tool based on data flow analysis. Wedge~\cite{bittau2008wedge} offers dynamic tracing components to analyze runtime memory access behavior. However, these tools still require developers to manually identify clear partition boundaries. To resolve this, other works have been proposed to support automated partitioning. Privtrans~\cite{brumley2004privtrans} uses static analysis to separate privileged and non-privileged processes. ProgramCutter~\cite{wu2013automatically} uses dynamic data dependency analysis to separate intra-process privileged code. PtrSplit~\cite{liu2017ptrsplit} supports general pointers by constructing a program dependence graph for each function. 
\sysname focuses on safeguarding keys and their propagation, using taint analysis to automate program partitioning. Specifically, after partitioning, CCAegis enables privilege-aware instrumentation (e.g., replacing key-related system calls) in \cref{subsec:instrumentation}.

\noindent \textbf{Intra-Process Isolation.}
Intra-process isolation~\cite{jin2022annotating,vahldiek2019erim} provides fine-grained protection for sensitive data.
Some works utilize existing isolation primitives. For example, 
libmpk~\cite{park2019libmpk} implements a secure and scalable isolation framework using the Intel Memory Protection Keys (MPK). 
\textsc{Cali}~\cite{bauer2021cali} employs \texttt{nsjail}~\cite{google2023nsjail} to isolate libraries. 
Other works construct their own isolation primitives. For example, PANIC~\cite{xu2023panic} utilizes Privileged Access Never (PAN) and Load Store Unit (LSU) to implement an access-control-based isolation mechanism. Capacity~\cite{dinh2023capacity} employs PAC and Memory Tagging Extension (MTE) to create a capability-based framework.
While these features effectively control memory access permissions within user space, they are not designed to protect against attacks from privileged software.

\noindent \textbf{CCA Enhancement.}
Although the hardware for CCA has not yet been released, several manufacturers are updating their software stacks. Arm has released a lightweight hypervisor for the realm world, the Realm Management Monitor (RMM)~\cite{arm2023rmm}. Samsung has implemented Islet~\cite{Samsung2023rmm}, a Rust-based RMM that supports confidential machine learning. Recently, some works have focused on verifying the TCB of CCA or applying CCA in various security contexts. The Verification Infrastructure for Armv9 (VIA)~\cite{li2022design} conducts formal verification of the RMM's security, while the Trusted Firmware Explorer (TFX)~\cite{fox2023verification} verifies the RMM and RME hardware implementation interface. Acai~\cite{sridhara2024acai} and CAGE~\cite{wang2024cage} enable CVMs in CCA to securely access accelerators. In addition to utilizing Realms, \textsc{Shelter}~\cite{zhang2023shelter} and RContainer~\cite{zhourcontainer} use GPTs to protect security-sensitive applications and containers in the normal world, respectively. However, none of these CCA-based systems provide function-level fine-grained isolation.

\section{Conclusion} \label{sec:conclusion}
\sysname is the first system to extend Arm CCA for fine-grained isolation. It decouples analysis and isolation, providing a generalized isolation capability for cryptographic programs. Utilizing static analysis, \sysname automates the partitioning of program components that require protection and employs three types of GPTs to safeguard cryptographic keys. Experiments demonstrate that \sysname has a more compact TCB, which effectively prevents key leakage from privileged and intra-process threats. Real-world applications such as \texttt{Nginx}, SSH connections, and popular cryptography libraries show only a modest overhead of 1.01$\times$ to 1.43$\times$ on our system.

\ifCLASSOPTIONcaptionsoff
  \newpage
\fi

\bibliographystyle{IEEEtran}
\bibliography{reference}

@inproceedings{durumeric2014matter,
  title={The matter of heartbleed},
  author={Durumeric, Zakir and Li, Frank and Kasten, James and Amann, Johanna and Beekman, Jethro and Payer, Mathias and Weaver, Nicolas and Adrian, David and Paxson, Vern and Bailey, Michael and others},
  booktitle={Proceedings of the 2014 conference on internet measurement conference},
  pages={475--488},
  year={2014}
}

@inproceedings{dautenhahn2015nested,
  title={Nested kernel: An operating system architecture for intra-kernel privilege separation},
  author={Dautenhahn, Nathan and Kasampalis, Theodoros and Dietz, Will and Criswell, John and Adve, Vikram},
  booktitle={Proceedings of the Twentieth International Conference on Architectural Support for Programming Languages and Operating Systems},
  pages={191--206},
  year={2015}
}

@inproceedings{gu2022hardware,
  title={A Hardware-Software co-design for efficient Intra-Enclave isolation},
  author={Gu, Jinyu and Zhu, Bojun and Li, Mingyu and Li, Wentai and Xia, Yubin and Chen, Haibo},
  booktitle={31st USENIX Security Symposium (USENIX Security 22)},
  pages={3129--3145},
  year={2022}
}

@inproceedings{park2019libmpk,
  title={libmpk: Software abstraction for intel memory protection keys (intel {MPK})},
  author={Park, Soyeon and Lee, Sangho and Xu, Wen and Moon, Hyungon and Kim, Taesoo},
  booktitle={2019 USENIX Annual Technical Conference (USENIX ATC 19)},
  pages={241--254},
  year={2019}
}

@inproceedings{jin2022annotating,
  title={Annotating, tracking, and protecting cryptographic secrets with CryptoMPK},
  author={Jin, Xuancheng and Xiao, Xuangan and Jia, Songlin and Gao, Wang and Gu, Dawu and Zhang, Hang and Ma, Siqi and Qian, Zhiyun and Li, Juanru},
  booktitle={2022 IEEE Symposium on Security and Privacy (SP)},
  pages={650--665},
  year={2022},
  organization={IEEE}
}

@inproceedings{vahldiek2019erim,
  title={{ERIM}: Secure, Efficient In-process Isolation with Protection Keys ({MPK})},
  author={Vahldiek-Oberwagner, Anjo and Elnikety, Eslam and Duarte, Nuno O and Sammler, Michael and Druschel, Peter and Garg, Deepak},
  booktitle={28th USENIX Security Symposium (USENIX Security 19)},
  pages={1221--1238},
  year={2019}
}

@inproceedings{xu2023panic,
  title={PANIC: PAN-assisted Intra-process Memory Isolation on {ARM}},
  author={Xu, Jiali and Xie, Mengyao and Wu, Chenggang and Zhang, Yinqian and Li, Qijing and Huang, Xuan and Lai, Yuanming and Kang, Yan and Wang, Wei and Wei, Qiang and others},
  booktitle={Proceedings of the 2023 ACM SIGSAC Conference on Computer and Communications Security},
  pages={919--933},
  year={2023}
}

@inproceedings{dinh2023capacity,
  title={Capacity: Cryptographically-Enforced In-Process Capabilities for Modern ARM Architectures},
  author={Dinh Duy, Kha and Cho, Kyuwon and Noh, Taehyun and Lee, Hojoon},
  booktitle={Proceedings of the 2023 ACM SIGSAC Conference on Computer and Communications Security},
  pages={874--888},
  year={2023}
}

@article{alves2004trustzone,
  title={Trustzone: Integrated hardware and software security},
  author={Alves, Tiago},
  journal={Information Quarterly},
  volume={3},
  pages={18--24},
  year={2004}
}

@misc{arm2023cca,
  title={Confidential Compute Architecture},
  author={Arm},
  year={2023},
  note={\url{https://www.arm.com/architecture/security-features/arm-confidential-compute-architecture}}
}

@inproceedings{zhang2023shelter,
  title={SHELTER: Extending Arm CCA with Isolation in User Space},
  author={Zhang, Yiming and Hu, Yuxin and Ning, Zhenyu and Zhang, Fengwei and Luo, Xiapu and Huang, Haoyang and Yan, Shoumeng and He, Zhengyu},
  booktitle={32nd USENIX Security Symposium (USENIX Security’23)},
  year={2023}
}

@inproceedings{machiry2017dr,
  title={{DR}.{CHECKER}: A soundy analysis for linux kernel drivers},
  author={Machiry, Aravind and Spensky, Chad and Corina, Jake and Stephens, Nick and Kruegel, Christopher and Vigna, Giovanni},
  booktitle={26th USENIX Security Symposium (USENIX Security 17)},
  pages={1007--1024},
  year={2017}
}

@inproceedings{yun2019ginseng,
  title={Ginseng: Keeping Secrets in Registers When You Distrust the Operating System.},
  author={Yun, Min Hong and Zhong, Lin},
  booktitle={NDSS},
  year={2019}
}

@misc{arm2023fvp,
  title={Fixed Virtual Platforms},
  author={Arm},
  year={2023},
  note={\url{https://developer.arm.com/downloads/-/arm-ecosystem-models}}
}

@inproceedings{lee2020off,
  title={An {Off-Chip} attack on hardware enclaves via the memory bus},
  author={Lee, Dayeol and Jung, Dongha and Fang, Ian T and Tsai, Chia-Che and Popa, Raluca Ada},
  booktitle={29th USENIX Security Symposium (USENIX Security 20)},
  year={2020}
}

@inproceedings{yitbarek2017cold,
  title={Cold boot attacks are still hot: Security analysis of memory scramblers in modern processors},
  author={Yitbarek, Salessawi Ferede and Aga, Misiker Tadesse and Das, Reetuparna and Austin, Todd},
  booktitle={2017 IEEE International Symposium on High Performance Computer Architecture (HPCA)},
  pages={313--324},
  year={2017},
  organization={IEEE}
}

@article{kim2014flipping,
  title={Flipping bits in memory without accessing them: An experimental study of DRAM disturbance errors},
  author={Kim, Yoongu and Daly, Ross and Kim, Jeremie and Fallin, Chris and Lee, Ji Hye and Lee, Donghyuk and Wilkerson, Chris and Lai, Konrad and Mutlu, Onur},
  journal={ACM SIGARCH Computer Architecture News},
  volume={42},
  number={3},
  pages={361--372},
  year={2014},
  publisher={ACM New York, NY, USA}
}

@article{lipp2018meltdown,
  title={Meltdown},
  author={Lipp, Moritz and Schwarz, Michael and Gruss, Daniel and Prescher, Thomas and Haas, Werner and Mangard, Stefan and Kocher, Paul and Genkin, Daniel and Yarom, Yuval and Hamburg, Mike},
  journal={arXiv preprint arXiv:1801.01207},
  year={2018}
}

@article{kocher2020spectre,
  title={Spectre attacks: Exploiting speculative execution},
  author={Kocher, Paul and Horn, Jann and Fogh, Anders and Genkin, Daniel and Gruss, Daniel and Haas, Werner and Hamburg, Mike and Lipp, Moritz and Mangard, Stefan and Prescher, Thomas and others},
  journal={Communications of the ACM},
  volume={63},
  number={7},
  pages={93--101},
  year={2020},
  publisher={ACM New York, NY, USA}
}

@inproceedings{shacham2007geometry,
  title={The geometry of innocent flesh on the bone: Return-into-libc without function calls (on the x86)},
  author={Shacham, Hovav},
  booktitle={Proceedings of the 14th ACM conference on Computer and communications security},
  pages={552--561},
  year={2007}
}

@misc{intel2023tdx,
  title={Intel Trust Domain Extensions (Intel {TDX})},
  author={Intel},
  year={2023},
  note={\url{https://www.intel.com/content/www/us/en/developer/articles/technical/intel-trust-domain-extensions.html}}
}

@misc{amd2023sev,
  title={Amd Secure Encrypted Virtualization ({SEV})},
  author={AMD},
  year={2023},
  note={\url{https://www.amd.com/en/developer/sev.html}}
}

@inproceedings{gharat2016flow,
  title={Flow-and context-sensitive points-to analysis using generalized points-to graphs},
  author={Gharat, Pritam M and Khedker, Uday P and Mycroft, Alan},
  booktitle={Static Analysis: 23rd International Symposium, SAS 2016, Edinburgh, UK, September 8-10, 2016, Proceedings 23},
  pages={212--236},
  year={2016},
  organization={Springer}
}

@inproceedings{sridhara2024acai,
  title={ACAI: Protecting Accelerator Execution with Arm Confidential Computing Architecture},
  author={Sridhara, Supraja and Bertschi, Andrin and Schl{\"u}ter, Benedict and Kuhne, Mark and Aliberti, Fabio and Shinde, Shweta},
  booktitle={33rd USENIX Security Symposium (USENIX Security 24)},
  pages={3423--3440},
  year={2024}
}

@misc{arm2023rmm,
  title={{TF-RMM}: An implementation of Arm-CCA RMM},
  author={Trusted Firmware},
  year={2023},
  note={\url{https://github.com/TF-RMM/tf-rmm}}
}

@misc{Samsung2023rmm,
  title={ISLET: An on-device confidential computing framework},
  author={Samsung},
  year={2023},
  note={\url{https://github.com/islet-project/islet}}
}

@article{fox2023verification,
  title={A Verification Methodology for the Arm{\textregistered} Confidential Computing Architecture: From a Secure Specification to Safe Implementations},
  author={Fox, Anthony CJ and Stockwell, Gareth and Xiong, Shale and Becker, Hanno and Mulligan, Dominic P and Petri, Gustavo and Chong, Nathan},
  journal={Proceedings of the ACM on Programming Languages},
  volume={7},
  number={OOPSLA1},
  pages={376--405},
  year={2023},
  publisher={ACM New York, NY, USA}
}

@inproceedings{li2022design,
  title={Design and verification of the arm confidential compute architecture},
  author={Li, Xupeng and Li, Xuheng and Dall, Christoffer and Gu, Ronghui and Nieh, Jason and Sait, Yousuf and Stockwell, Gareth},
  booktitle={16th USENIX Symposium on Operating Systems Design and Implementation (OSDI 22)},
  pages={465--484},
  year={2022}
}

@misc{intel2023sgx,
  title={Intel® Software Guard Extensions (Intel® {SGX})},
  author={Intel},
  year={2023},
  note={\url{https://www.intel.com/content/www/us/en/architecture-and-technology/software-guard-extensions.html}}
}

@article{checkoway2013iago,
  title={Iago attacks: Why the system call API is a bad untrusted RPC interface},
  author={Checkoway, Stephen and Shacham, Hovav},
  journal={ACM SIGARCH Computer Architecture News},
  volume={41},
  number={1},
  pages={253--264},
  year={2013},
  publisher={ACM New York, NY, USA}
}

@inproceedings{cui2021emilia,
  title={Emilia: Catching Iago in Legacy Code.},
  author={Cui, Rongzhen and Zhao, Lianying and Lie, David},
  booktitle={NDSS},
  year={2021}
}

@inproceedings{brasser2019sanctuary,
  title={SANCTUARY: ARMing TrustZone with User-space Enclaves.},
  author={Brasser, Ferdinand and Gens, David and Jauernig, Patrick and Sadeghi, Ahmad-Reza and Stapf, Emmanuel},
  booktitle={NDSS},
  year={2019}
}

@misc{google2023nsjail,
  title={nsjail: a light-weight process isolation tool, making use of Linux namespaces and seccomp-bpf syscall filters},
  author={Google},
  year={2023},
  note={\url{https://nsjail.dev/}}
}

@article{grech2017p,
  title={P/taint: Unified points-to and taint analysis},
  author={Grech, Neville and Smaragdakis, Yannis},
  journal={Proceedings of the ACM on Programming Languages},
  volume={1},
  number={OOPSLA},
  pages={1--28},
  year={2017},
  publisher={ACM New York, NY, USA}
}

@inproceedings{lin2022dirtycred,
  title={DirtyCred: Escalating Privilege in Linux Kernel},
  author={Lin, Zhenpeng and Wu, Yuhang and Xing, Xinyu},
  booktitle={Proceedings of the 2022 ACM SIGSAC Conference on Computer and Communications Security},
  pages={1963--1976},
  year={2022}
}

@inproceedings{wang2024cage,
title={CAGE: Complementing Arm CCA with GPU Extensions},
author={Wang, Chenxu and Zhang, Fengwei and Deng, Yunjie and Leach, Kevin and Cao, Jiannong and Ning, Zhenyu and Yan, Shoumeng and He, Zhengyu},
booktitle={Proceedings of the 31st Annual Network and Distributed System Security Symposium},
year={2024}
}

@inproceedings{lee2020keystone,
  title={Keystone: An open framework for architecting trusted execution environments},
  author={Lee, Dayeol and Kohlbrenner, David and Shinde, Shweta and Asanovi{\'c}, Krste and Song, Dawn},
  booktitle={Proceedings of the Fifteenth European Conference on Computer Systems},
  pages={1--16},
  year={2020}
}

@inproceedings{feng2021scalable,
  title={Scalable memory protection in the {PENGLAI} enclave},
  author={Feng, Erhu and Lu, Xu and Du, Dong and Yang, Bicheng and Jiang, Xueqiang and Xia, Yubin and Zang, Binyu and Chen, Haibo},
  booktitle={15th {USENIX} Symposium on Operating Systems Design and Implementation ({OSDI} 21)},
  pages={275--294},
  year={2021}
}

@misc{cloc,
  title={cloc: Count Lines of Code},
  author={Danial, Al},
  year={2024},
  note={\url{https://github.com/AlDanial/cloc}}
}

@misc{arm2023tfrmm,
  title={Reference implementation of Arm-CCA RMM specification},
  author={Arm},
  year={2023},
  note={\url{https://github.com/TF-RMM/tf-rmm/releases/tag/tf-rmm-v0.4.0}}
}

@misc{arm2024ccaguest,
  title={Guest Arm CCA Linux branches},
  author={Arm},
  year={2024},
  note={\url{https://gitlab.arm.com/linux-arm/linux-cca/-/tree/cca-guest/v2}}
}

@inproceedings{zhou2014armlock,
  title={Armlock: Hardware-based fault isolation for arm},
  author={Zhou, Yajin and Wang, Xiaoguang and Chen, Yue and Wang, Zhi},
  booktitle={Proceedings of the 2014 ACM SIGSAC conference on computer and communications security},
  pages={558--569},
  year={2014}
}

@misc{arm2021rme,
  title={Learn the architecture - {Realm} Management Extension},
  author={Arm},
  year={2021},
  note={\url{https://developer.arm.com/documentation/den0126/latest/}}
}

@inproceedings{bratus2008toctou,
  title={TOCTOU, traps, and trusted computing},
  author={Bratus, Sergey and D’Cunha, Nihal and Sparks, Evan and Smith, Sean W},
  booktitle={International Conference on Trusted Computing},
  pages={14--32},
  year={2008},
  organization={Springer}
}

@inproceedings{shinde2020besfs,
  title={{BesFS}: A {POSIX} Filesystem for Enclaves with a Mechanized Safety Proof},
  author={Shinde, Shweta and Wang, Shengyi and Yuan, Pinghai and Hobor, Aquinas and Roychoudhury, Abhik and Saxena, Prateek},
  booktitle={29th USENIX Security Symposium (USENIX Security 20)},
  pages={523--540},
  year={2020}
}

@inproceedings{zhang2016case,
  title={Case: Cache-assisted secure execution on arm processors},
  author={Zhang, Ning and Sun, Kun and Lou, Wenjing and Hou, Y Thomas},
  booktitle={2016 IEEE Symposium on Security and Privacy (SP)},
  pages={72--90},
  year={2016},
  organization={IEEE}
}

@inproceedings{bahmani2021cure,
  title={{CURE}: A security architecture with {CUstomizable} and resilient enclaves},
  author={Bahmani, Raad and Brasser, Ferdinand and Dessouky, Ghada and Jauernig, Patrick and Klimmek, Matthias and Sadeghi, Ahmad-Reza and Stapf, Emmanuel},
  booktitle={30th USENIX Security Symposium (USENIX Security 21)},
  pages={1073--1090},
  year={2021}
}

@inproceedings{costan2016sanctum,
  title={Sanctum: Minimal hardware extensions for strong software isolation},
  author={Costan, Victor and Lebedev, Ilia and Devadas, Srinivas},
  booktitle={25th USENIX Security Symposium (USENIX Security 16)},
  pages={857--874},
  year={2016}
}

@misc{keystone2023eyrie,
  title={Eyrie enclave runtime kernel},
  author={Keystone},
  year={2023},
  note={\url{https://github.com/keystone-enclave/keystone-runtime}}
}

@inproceedings{haller2013dowsing,
  title={Dowsing for {Overflows}: A Guided Fuzzer to Find Buffer Boundary Violations},
  author={Haller, Istvan and Slowinska, Asia and Neugschwandtner, Matthias and Bos, Herbert},
  booktitle={22nd USENIX Security Symposium (USENIX Security 13)},
  pages={49--64},
  year={2013}
}

@article{ramalingam1994undecidability,
  title={The undecidability of aliasing},
  author={Ramalingam, Ganesan},
  journal={ACM Transactions on Programming Languages and Systems (TOPLAS)},
  volume={16},
  number={5},
  pages={1467--1471},
  year={1994},
  publisher={ACM New York, NY, USA}
}

@inproceedings{hardekopf2011flow,
  title={Flow-sensitive pointer analysis for millions of lines of code},
  author={Hardekopf, Ben and Lin, Calvin},
  booktitle={International Symposium on Code Generation and Optimization (CGO 2011)},
  pages={289--298},
  year={2011},
  organization={IEEE}
}

@inproceedings{hanpage2024pop,
  title={Page-Oriented Programming: Subverting Control-Flow Integrity of Commodity Operating System Kernels with Non-Writable Code Pages},
  author={Han, Seunghun and Kim, Seong-Joong and Shin, Wook and Kim, Byung Joon and Ryou, Jae-Cheol},
  booktitle={33nd USENIX Security Symposium (USENIX Security’24)},
  year={2024}
}

@inproceedings{chen2016shreds,
  title={Shreds: Fine-grained execution units with private memory},
  author={Chen, Yaohui and Reymondjohnson, Sebassujeen and Sun, Zhichuang and Lu, Long},
  booktitle={2016 IEEE Symposium on Security and Privacy (SP)},
  pages={56--71},
  year={2016},
  organization={IEEE}
}

@inproceedings{brumley2004privtrans,
  title={Privtrans: Automatically partitioning programs for privilege separation},
  author={Brumley, David and Song, Dawn},
  booktitle={USENIX Security Symposium},
  volume={57},
  number={72},
  year={2004}
}

@inproceedings{bittau2008wedge,
  title={Wedge: splitting applications into reduced-privilege compartments},
  author={Bittau, Andrea and Marchenko, Petr and Handley, Mark and Karp, Brad},
  booktitle={Proceedings of the 5th USENIX Symposium on Networked Systems Design and Implementation},
  pages={309--322},
  year={2008}
}

@inproceedings{wu2013automatically,
  title={Automatically partition software into least privilege components using dynamic data dependency analysis},
  author={Wu, Yongzheng and Sun, Jun and Liu, Yang and Dong, Jin Song},
  booktitle={2013 28th IEEE/ACM International Conference on Automated Software Engineering (ASE)},
  pages={323--333},
  year={2013},
  organization={IEEE}
}

@inproceedings{liu2017ptrsplit,
  title={Ptrsplit: Supporting general pointers in automatic program partitioning},
  author={Liu, Shen and Tan, Gang and Jaeger, Trent},
  booktitle={Proceedings of the 2017 ACM SIGSAC Conference on Computer and Communications Security},
  pages={2359--2371},
  year={2017}
}

@inproceedings{bauer2021cali,
  title={Cali: Compiler-assisted library isolation},
  author={Bauer, Markus and Rossow, Christian},
  booktitle={Proceedings of the 2021 ACM Asia Conference on Computer and Communications Security},
  pages={550--564},
  year={2021}
}

@inproceedings{gudka2015clean,
  title={Clean application compartmentalization with SOAAP},
  author={Gudka, Khilan and Watson, Robert NM and Anderson, Jonathan and Chisnall, David and Davis, Brooks and Laurie, Ben and Marinos, Ilias and Neumann, Peter G and Richardson, Alex},
  booktitle={Proceedings of the 22nd ACM SIGSAC Conference on Computer and Communications Security},
  pages={1016--1031},
  year={2015}
}

@inproceedings{liu2015thwarting,
  title={Thwarting memory disclosure with efficient hypervisor-enforced intra-domain isolation},
  author={Liu, Yutao and Zhou, Tianyu and Chen, Kexin and Chen, Haibo and Xia, Yubin},
  booktitle={Proceedings of the 22nd ACM SIGSAC Conference on Computer and Communications Security},
  pages={1607--1619},
  year={2015}
}

@inproceedings{lind2017glamdring,
  title={Glamdring: Automatic application partitioning for intel {SGX}},
  author={Lind, Joshua and Priebe, Christian and Muthukumaran, Divya and O'Keeffe, Dan and Aublin, Pierre-Louis and Kelbert, Florian and Reiher, Tobias and Goltzsche, David and Eyers, David and Kapitza, R{\"u}diger and others},
  booktitle={2017 USENIX Annual Technical Conference (USENIX ATC 17)},
  pages={285--298},
  year={2017}
}

@inproceedings{zhourcontainer,
  title={RContainer: A Secure Container Architecture through Extending ARM CCA Hardware Primitives},
  author={Zhou, Qihang and Cao, Wenzhuo and Jia, Xiaoqi and Liu, Peng and Zhang, Shengzhi and Chen, Jiayun and Xu, Shaowen and Song, Zhenyu},
  booktitle={NDSS},
  year={2025}
}

@inproceedings{park2020nested,
  title={Nested enclave: Supporting fine-grained hierarchical isolation with sgx},
  author={Park, Joongun and Kang, Naegyeong and Kim, Taehoon and Kwon, Youngjin and Huh, Jaehyuk},
  booktitle={2020 ACM/IEEE 47th Annual International Symposium on Computer Architecture (ISCA)},
  pages={776--789},
  year={2020},
  organization={IEEE}
}

@misc{arm2024tfa,
  title={Trusted Firmware-A Documentation},
  author={Arm},
  year={2024},
  note={\url{https://trustedfirmware-a.readthedocs.io/en/latest/}}
}
\vfill

\begin{IEEEbiography} [{\includegraphics[width=1in,height=1.25in,clip,keepaspectratio]{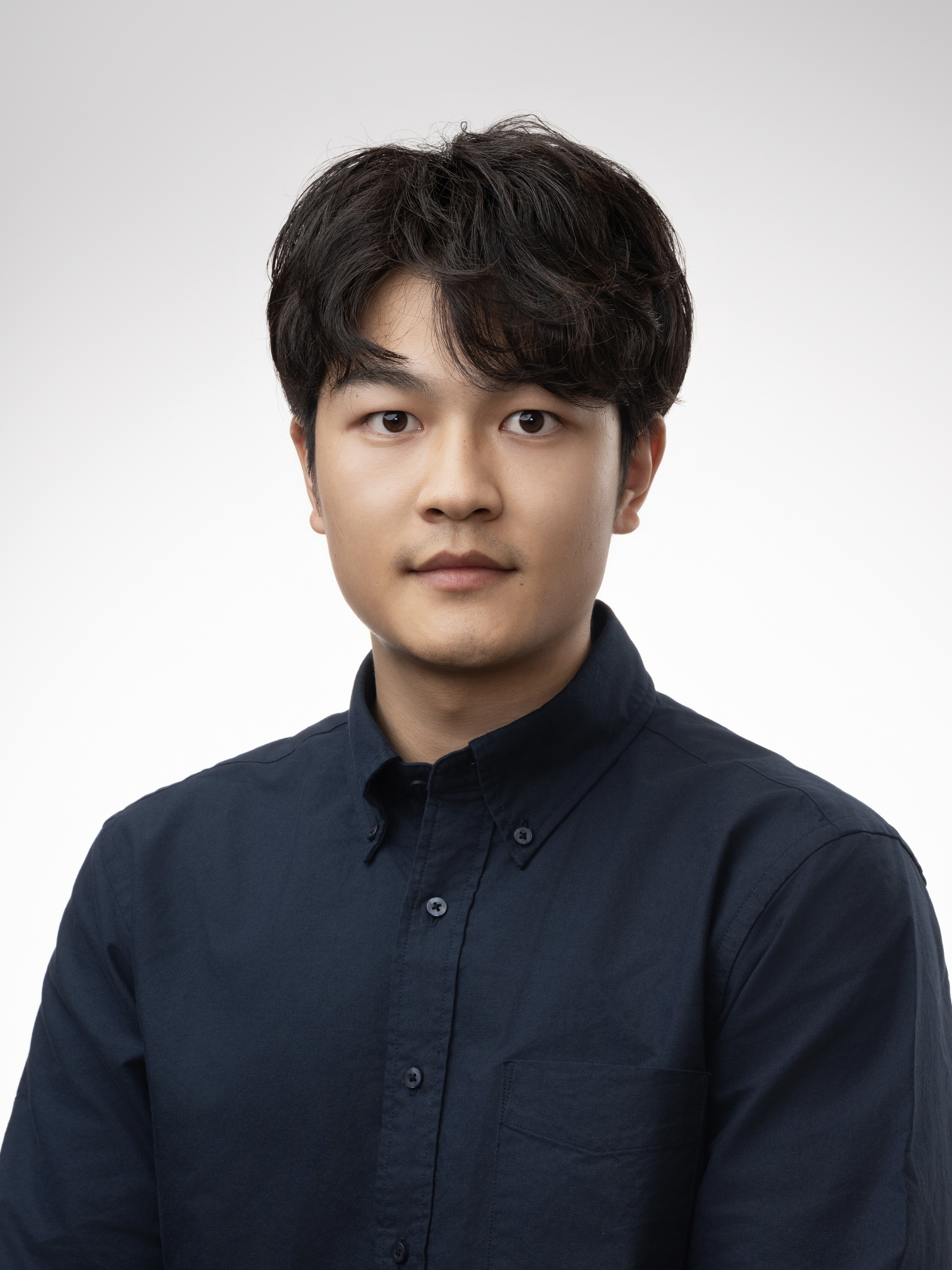}}]
{Shiqi Liu} 
received the B.S. degree in Cyber Science and Engineering from the University of International Relations, Beijing, China, in 2022, and the M.S. degree in Cyber Science and Engineering from Huazhong University of Science and Technology, Wuhan, China, in 2025. He is currently a Ph.D. student with the Center for Secure Information Systems at George Mason University, Fairfax, VA, USA. His research interests include system security and software security.
\end{IEEEbiography}

\begin{IEEEbiography} [{\includegraphics[width=1in,height=1.25in,clip,keepaspectratio]{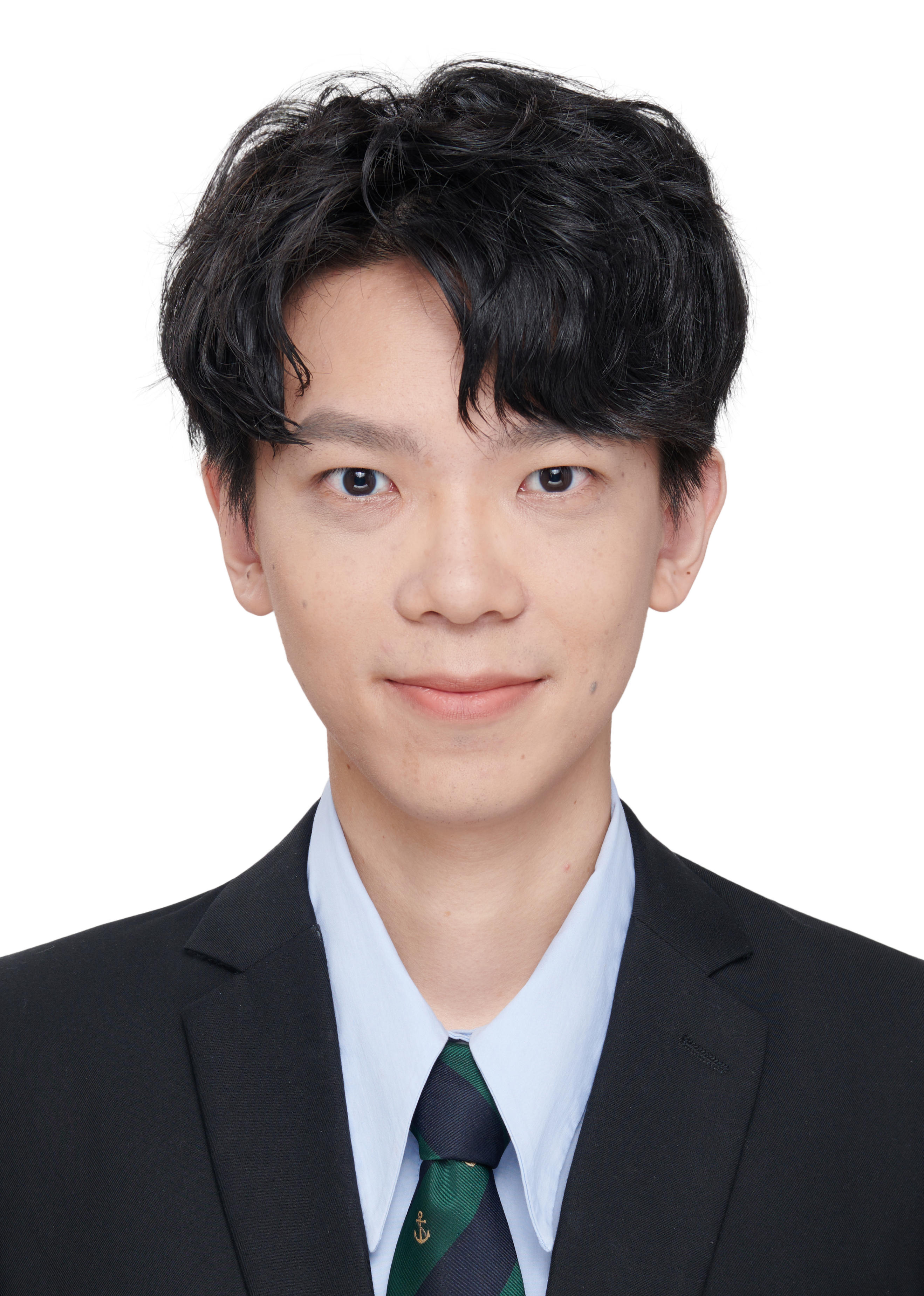}}]
{Zhouqi Jiang}
received the B.S. degree in Cyber Science and Engineering from Huazhong University of Science and Technology, Wuhan, China, in 2022, and the M.S. degree in Cyber Science and Engineering from the same university in 2025. He is currently pursuing the Ph.D. degree with the Institute of Software, Chinese Academy of Sciences (ISCAS), and the University of Chinese Academy of Sciences, Beijing, China. His research interests include secure firmware and system security.
\end{IEEEbiography} 

\begin{IEEEbiography} [{\includegraphics[width=1in,height=1.25in,clip,keepaspectratio]{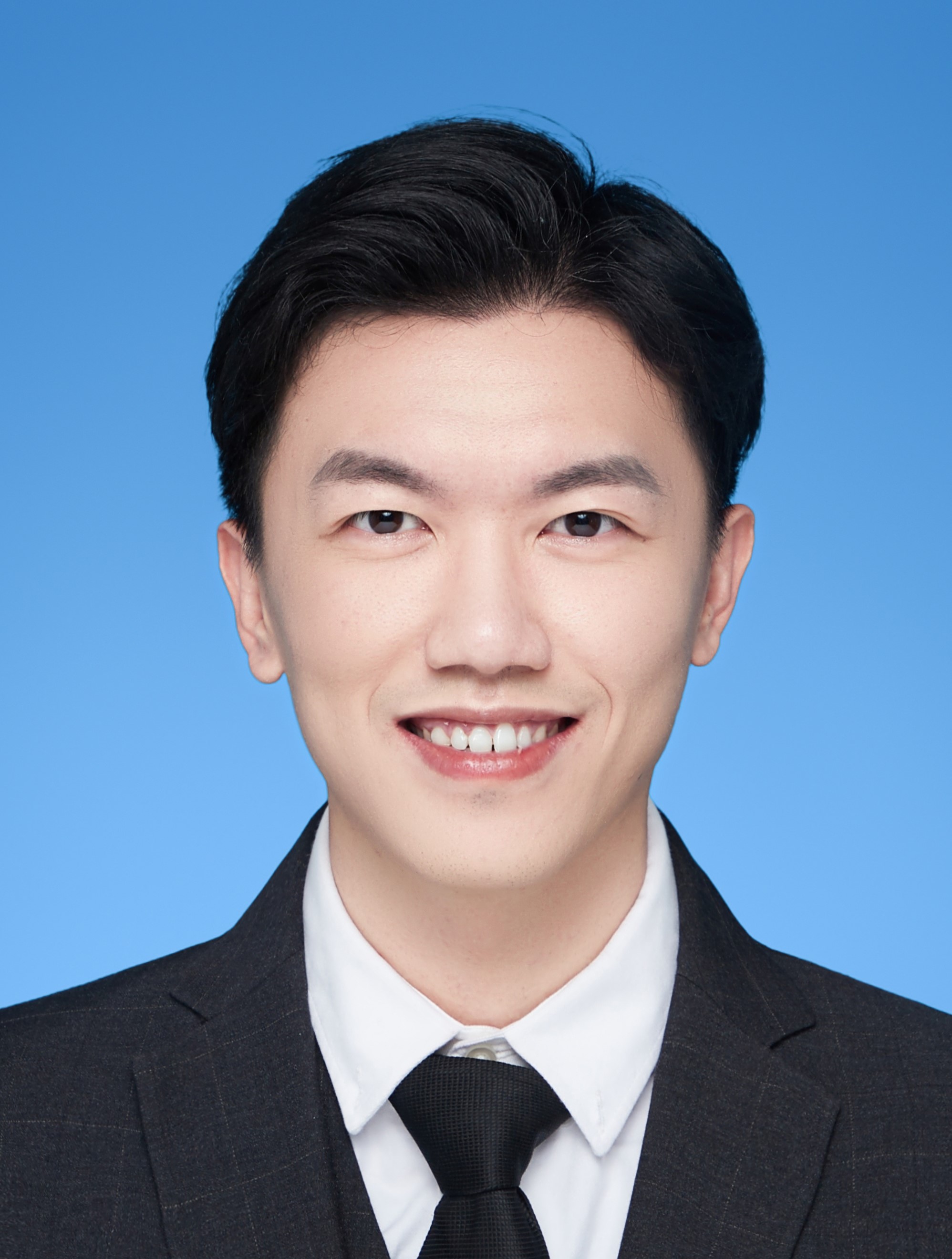}}]
{Jie Wang}
received his Ph.D. degree from the University of Chinese Academy of Sciences (UCAS) in 2021. He was a visiting scholar at George Mason University and is currently an Associate Professor at the School of Cyber Science and Engineering, Huazhong University of Science and Technology. His research interests include system security, trusted computing, and trusted AI accelerators.
\end{IEEEbiography} 

\begin{IEEEbiography} [{\includegraphics[width=1in,height=1.25in,clip,keepaspectratio]{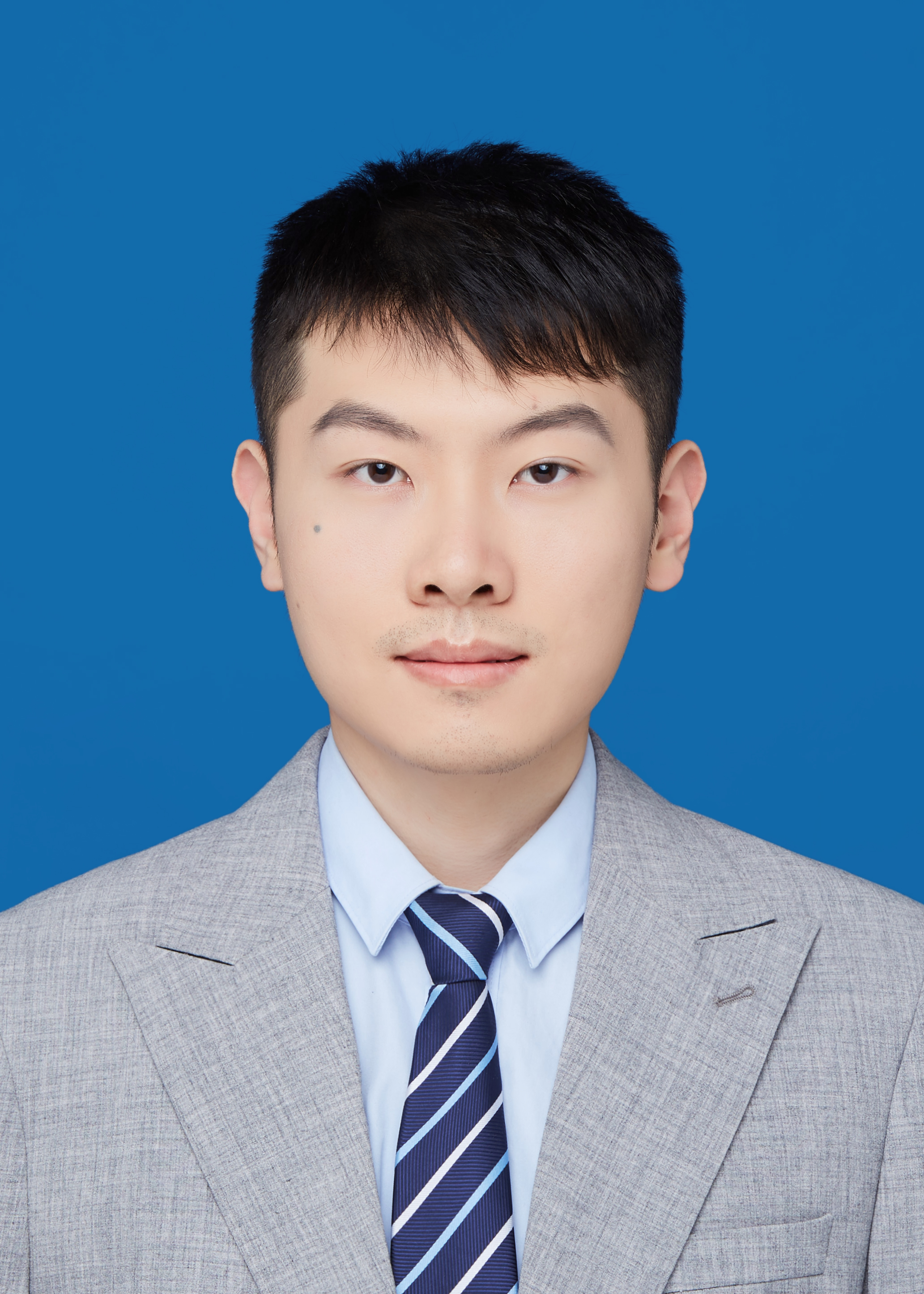}}]
{Wei Zhou}
received his Ph.D. degree from the University of Chinese Academy of Sciences (UCAS) in 2021. He was a visiting scholar at Pennsylvania State University and is currently an Associate Professor at the School of Cyber Science and Engineering, Huazhong University of Science and Technology. His research interests include system security, trusted computing, and IoT system security.
\end{IEEEbiography}

\begin{IEEEbiography} [{\includegraphics[width=1in,height=1.25in,clip,keepaspectratio]{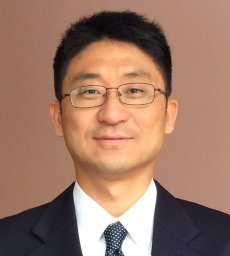}}]
{Kun Sun} (Member, IEEE)
received the Ph.D. degree in computer science from North Carolina State University. He has more than 15 years working experience in both industry and academia, and serves as the Director of the Sun Security Laboratory (SunLab) and the Associate Director of the Center for Secure Information Systems (CSIS). He has published more than 100 peer-reviewed conference and journal articles. His research focuses on systems and network security.
\end{IEEEbiography}

\begin{IEEEbiography} [{\includegraphics[width=1in,height=1.25in,clip,keepaspectratio]{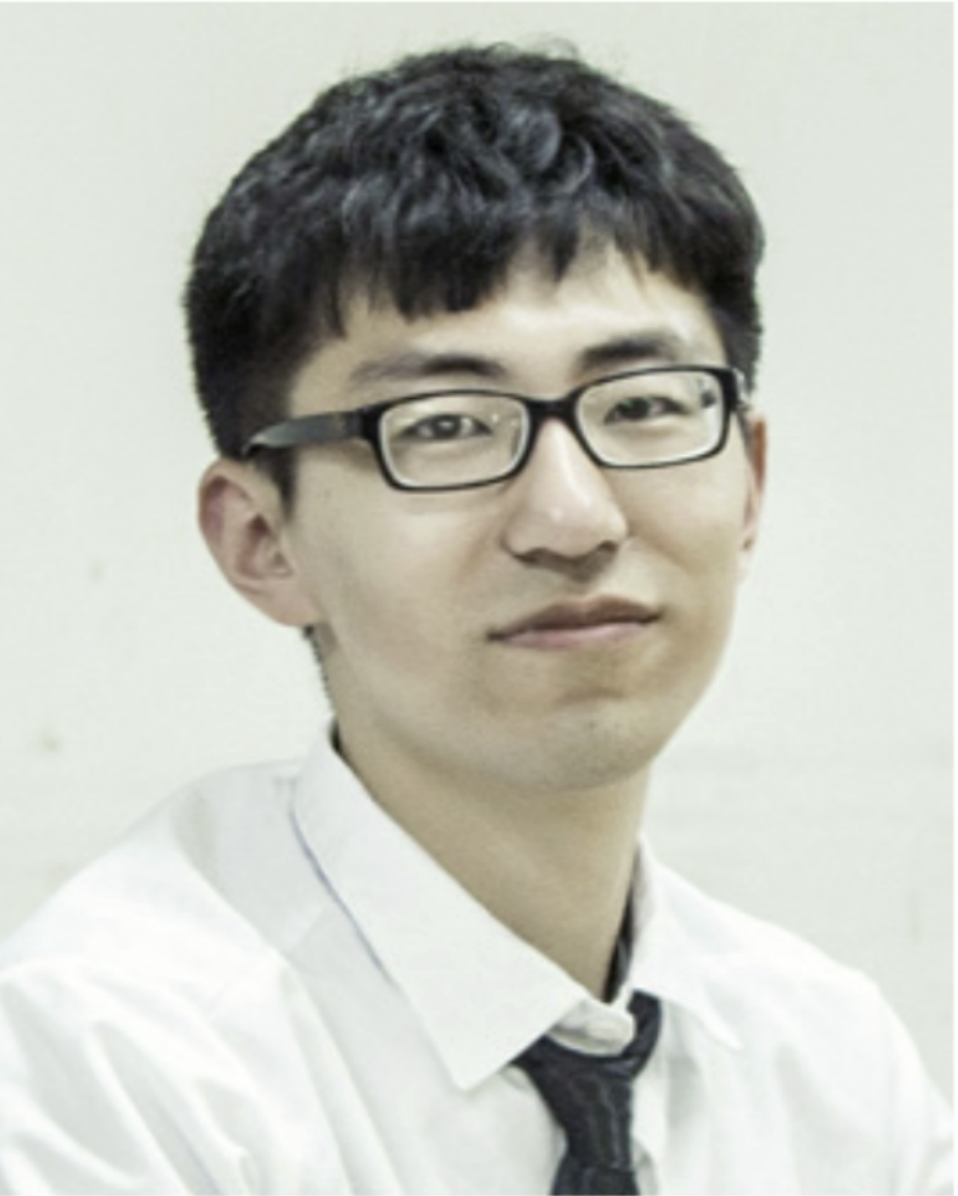}}]
{Zhaohui Chen}
received the Ph.D. degree in computer science and technology from the University of Chinese Academy of Sciences (UCAS), Beijing, China, in 2022. He is currently working with DAMO Academy, Alibaba Group, Hangzhou, China. His research interests include applied cryptography, hardware security, and homomorphic encryption.
\end{IEEEbiography}

\begin{IEEEbiography} [{\includegraphics[width=1in,height=1.25in,clip,keepaspectratio]{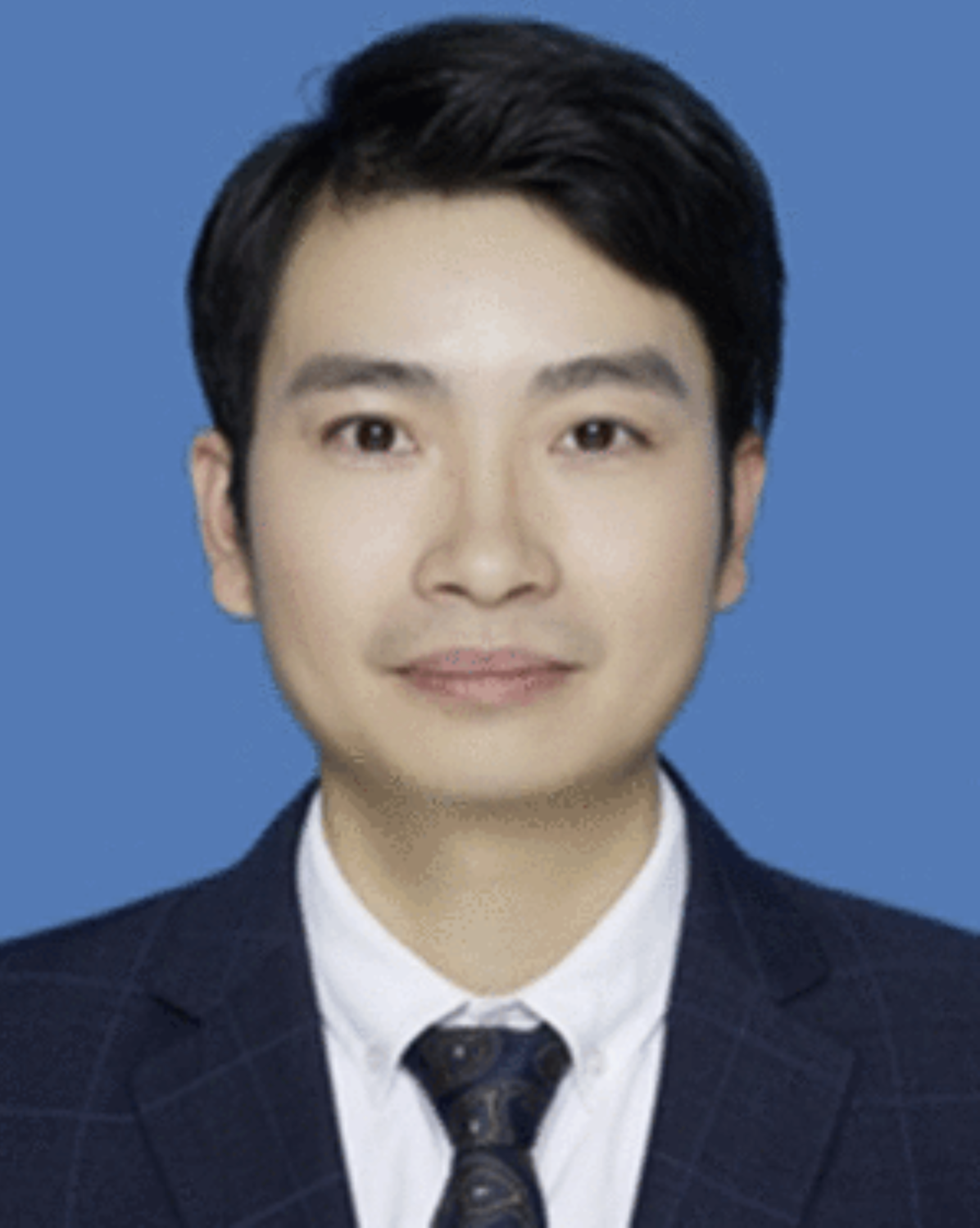}}]
{Yulai Xie} (Member, IEEE) 
received the B.E. and Ph.D. degrees in computer science from the Huazhong University of Science and Technology (HUST), China, in 2007 and 2013, respectively. He is currently a Professor with the School of Cyber Science and Engineering in HUST, China. He was a Visiting Scholar with the University of California, Santa Cruz, CA, USA, in 2010, and a Visiting Scholar with the Chinese University of Hong Kong, Hong Kong, in 2015. His research interests mainly include cloud storage and virtualization, digital provenance, intrusion detection, machine learning, and computer architecture.
\end{IEEEbiography}

\end{document}